\documentclass{lhep}       
\journal{LHEP}
\vol{xx}
\jyear{2018}
\pages{xxx} 
\received{xx January 2018}
\published{xx March 2018}
\def\be{\begin{equation}}
\def\ee{\end{equation}}
\def\bea{\begin{eqnarray}}
\def\eea{\end{eqnarray}}

\usepackage{bbm}
\usepackage{subcaption}
\usepackage{mathtools}
\usepackage{tikz}
\usepackage{pgfplots}
\pgfplotsset{compat=newest}
\usetikzlibrary{arrows,shapes}
\usetikzlibrary{trees}
\usetikzlibrary{matrix,arrows} 				
\usetikzlibrary{positioning}				
\usetikzlibrary{calc,through}			
\usetikzlibrary{decorations.pathreplacing} 
\usepackage{pgffor}							
\usetikzlibrary{decorations.pathmorphing}	
\usetikzlibrary{decorations.markings}

\tikzstyle{block} = [draw, rectangle, minimum height=3em, minimum width=6em]
\begin{document}
\title{Normalization of the Ground State of the Supersymmetric \\Harmonic Oscillator}
\author{Ahmed Ayad}
\address{School of Physics, University of the Witwatersrand, Private Bag 3, WITS-2050, Johannesburg, South Africa}

\begin{abstract}
Supersymmetry plays a main role in all current thinking about superstring theory. Indeed, many remarkable properties of string theory have been explained using supersymmetry as a tool. So far, there has been no unbroken supersymmetry observed in nature, and if nature is described by supersymmetry, it must be broken. Supersymmetry may be broken spontaneously at any order of perturbation theory or dynamically due to nonperturbative effects. To examine the methods of supersymmetry breaking, special attention is given to discuss the normalization of the ground state of the supersymmetric harmonic oscillator. This study explains that perturbation theory gives incorrect results for both the ground-state wave function and the energy spectrum and it fails to give an explanation to the supersymmetry breaking.
\end{abstract}
\maketitle

\section{Introduction}

Supersymmetry, often abbreviated SUSY, was first introduced in 1971 by Gel’fand and Likhtman, Raymond, and Neveu and Schwartz, and later, it was rediscovered by other groups \cite{ engbrant2012supersymmetric, Wellman}. Supersymmetric quantum mechanics was developed by Witten in 1982 \cite{witten1981dynamical}, as a toy model to test the breaking of supersymmetry. In general, supersymmetry plays a main role in the modern understanding of theoretical physics, in particular, quantum field theories, gravity theories, and string theories \cite{dumitrescu2013topics}.

Supersymmetry is a symmetry that connects particles of integer spins (bosons) and particles of half-integer spins (fermions) \cite{bertolini2011lectures}. In supersymmetry, it is possible to introduce operators that change bosons which are commuting variables into fermions which are anticommuting variables and vice versa \cite{cooper1995supersymmetry}. Supersymmetric theories are theories which are invariant under those transformations \cite{batzing2013lecture}. The supersymmetry algebra involves commutators as well as anticommutators \cite{krippendorf2010cambridge}. 

In this study, the main mathematical structure which involves supersymmetric quantum mechanics is derived starting with explaining the basic idea of supersymmetry and followed by introducing the necessary framework to make a supersymmetric theory. We derive the basic formulation of supersymmetric quantum mechanics starting with introducing the concepts of supercharges and superalgebra. We show that if there is a supersymmetric state, it is the zero-energy ground state. If such a state exists, the supersymmetry is unbroken; otherwise, it is broken. So far, there has been no unbroken supersymmetry observed in nature, and if nature is described by supersymmetry, it must be broken. In fact, supersymmetry may be broken spontaneously at any order of perturbation theory or dynamically due to nonperturbative effects. To examine the methods of supersymmetry breaking, we study the normalization of the ground state of the supersymmetric harmonic oscillator and calculate the corrections to the ground-state energy using perturbation theory. We found out that no perturbation effect can lead to supersymmetry breaking and it must be due to the nonperturbative effects.

This work is structured as follows. After a brief introduction, we study the problem of a harmonic oscillator with fermionic as well as bosonic fields using the usual quantum mechanical operators method. Then, we consider supersymmetric classical mechanics and study generalized classical Poisson brackets to Dirac brackets and quantization rules in order to introduce the superalgebra. Furthermore, the formalism of supersymmetry in quantum mechanics is introduced. Subsequently, we give a general background in the concepts and methods of supersymmetric quantum mechanics. This followed by providing a more specific study of the ground state of the supersymmetric harmonic oscillator. Finally, our conclusion is given, and for completeness, some needed derivations are broadly outlined in the appendix.

\section{Harmonic Oscillator} \label{sec2.2}

Supersymmetric theories necessarily include both bosons and fermions, so  a system that includes both bosonic and fermionic degrees of freedom is considered here. The supersymmetry algebra is a mathematical formalism for describing the relation between bosons and fermions, and we will discuss this later in Section \ref{Sec2.4}. In quantum mechanics, bosons are integer spin particles which obey Bose-Einstein statistics, and the bosonic operators satisfy the usual commutation relations, while fermions are half-integer spin particles characterized by Fermi–Dirac statistics, obey the Pauli exclusion principle, and can be described by Grassmann variables, which are anticommuting objects; for review, see \cite{wasay2010supersymmetric, wasay2016supersymmetric, morgan2003supersymmetric,seiberg2003noncommutative}.
 
The simplest system that consists of a combination of bosonic and fermionic fields is the one-dimensional harmonic oscillator. There are many excellent reviews on this subject; we refer the reader to them for more and complete expositions \cite{Wellman,bagchi2000supersymmetry,hamdouni2005supersymmetry,nakahara2003geometry}. We begin by considering the Lagrangian formulation for the bosonic and fermionic harmonic oscillator:
\begin{equation}\label{2.4}
   \mathit{L} = \frac{1}{2} \left( \dot{q}^2 + i \psi_{\alpha} \delta^{\alpha \beta} \dot{\psi}_{\beta} + \mathit{F}^2 \right) + \omega \left( i \psi_1 \psi_2 + q \mathit{F} \right) \:,
  \end{equation}    
where $q(t)$ and $\mathit{F}(t)$ are real bosonic fields, $\psi_{\alpha}(t)$ (with $\alpha \in \{1,2\}$) are real fermionic fields, $\omega$ is a real parameter, $\delta^{\alpha \beta}$ is the Kronecker delta, and we used the Einstein summation convention.

It is straightforward, using the classical equation of motion for $\mathit{F}+ \omega q =0$, to eliminate $\mathit{F}$ from Eq. (\ref{2.4}) and obtain the equivalent Lagrangian:
\begin{equation} \label{2.5}
\mathit{L} = \frac{1}{2} \left( \dot{q}^2 + i \psi_{\alpha} \delta^{\alpha \beta} \dot{\psi}_{\beta} - \omega^2 q^2 \right) + i \omega \psi_1 \psi_2 \:.
\end{equation} 
Then, using the Legendre transformation from the variables $\{q,\dot{q},\psi,\dot{\psi}\}$ to $\{q,p,\psi,\pi\}$, we obtain
  \begin{equation}\label{2.6}
  \mathit{H}=\dot{q}p+\psi_{\alpha} \delta^{\alpha \beta} \pi^{\beta} - \mathit{L} = \frac{1}{2} \left( p^2 + \omega^2 q^2 \right) - i \omega \psi_1 \psi_2 \:,
  \end{equation}
where $\pi^{\alpha}$ are the momenta conjugate to $\psi_{\alpha}$.

Later in this study, we shall see that the general properties of supersymmetry have to be clear if the Hamiltonian is built as a larger Hamiltonian consisting of two components, one representing the bosonic field and the other representing the fermionic field. Leaving this aside for the moment, to derive the supersymmetric Hamiltonian, let us define the bosonic valuable $\hat{q}$ and the bosonic momentum $\hat{p}$, respectively, as follows:
\begin{align} \label{2.7}
&&&& \hat{q} &= q \mathbbmtt{I}_2, &&  \hat{p} = - i \hbar \mathbbmtt{I}_2 \dfrac{\partial}{\partial q} \:. &&&&
\end{align}
On the other hand, the fermionic variables $\hat{\psi}_1$ and $\hat{\psi}_2$ and the fermionic momenta $\pi^{\alpha}$, respectively, have to be defined as
\begin{align} \label{2.8}
 &&&&&& \hat{\psi}_1 &= \sqrt{\frac{\hbar}{2}} \sigma_1, && \hat{\psi}_2 = \sqrt{\frac{\hbar}{2}} \sigma_2, &&&&&& \nonumber \\
&&&&&& \pi^{\alpha} &= \frac{\partial \mathit{L}}{\partial \dot{\psi}_{\alpha}}= - \frac{i}{2} \delta^{\alpha \beta} \psi_{\beta},  && \alpha, \beta \in \{1,2 \} \:. &&&&&&
  \end{align}
Hence, using Eqs. (\ref{2.7}, \ref{2.8}), we can rewrite the previous Hamiltonian (\ref{2.6}) in the following form:          
  \begin{equation} \label{2.9}
  \mathit{\hat{H}} = \frac{1}{2} \left(- \hbar^2 \dfrac{d^2}{dq^2} + \omega^2 q^2 \right) \mathbbmtt{I}_2 + \left( \frac{1}{2} \hbar \omega \right) \sigma_3 \:,
  \end{equation}
where $\mathbbmtt{I}_2$ is the $2 \times 2$ identity matrix and $\sigma_j$ are the Pauli matrices; that is,
\begin{equation}
 \resizebox{.42 \textwidth}{!} {$
\mathbbmtt{I}_2 = \left( \begin{matrix} 1 & 0 \\ 0 & 1 \end{matrix} \right), \quad
\sigma_1 = \left( \begin{matrix} 0 & 1 \\ 1 & 0 \end{matrix} \right), \quad
\sigma_2 = \left( \begin{matrix} 0 & -i \\ i & 0 \end{matrix} \right), \quad
\sigma_3 = \left( \begin{matrix} 1 & 0 \\ 0 & -1 \end{matrix} \right) \:.$}
\end{equation}
We want to obtain solutions of the time-independent Schr\"{o}dinger equation $\mathit{H}\Psi = \mathit{E} \Psi$; that is,
  \begin{equation}
  \frac{1}{2} \left(- \hbar^2 \dfrac{d^2}{dq^2} + \omega^2 q^2 \right) \mathbbmtt{I}_2 + \left( \frac{1}{2} \hbar \omega \right) \sigma_3 = \mathit{E} \Psi \:,
  \end{equation}
with the wave function $\Psi(x)$ constraint to satisfy appropriate boundary conditions.

As we mentioned, the supersymmetric Hamiltonian can be written as a summation of two terms representing the bosonic and the fermionic components of the Hamiltonian
  \begin{equation}
  \mathit{H} = \mathit{H}_B (p,q)+\mathit{H}_F(\psi) \:.
  \end{equation}
To this end, we define the lowering (annihilation) and raising (creation) operators for bosons $\hat{a}^{\dagger}$ and $\hat{a}$, respectively, they are 
  \begin{align} \label{2.13}
&&  \hat{a} &= \sqrt{\frac{1}{2 \hbar \omega}} \left( \omega \hat{q}+i\hat{p} \right), &
   \hat{a}^{\dagger} &= \sqrt{\frac{1}{2 \hbar \omega}} \left( \omega \hat{q}-i\hat{p} \right) \:.&& 
  \end{align}
Later, in Section \ref{Sec2.4}, we will show that the bosonic operators, $\hat{q}$ and $\hat{p}$, satisfy the usual canonical quantum commutation condition $[\hat{q},\hat{p}]= i\hbar$. Taking this fact into account, we can find that the bosonic ladder operators $\hat{a}$ and $\hat{a}^{\dagger}$ satisfy the following commutation relations
  \begin{equation} \label{2.14}
 [\hat{a},\hat{a}^{\dagger}] =1, \quad \quad [\hat{a},\hat{a} ]  = [\hat{a}^{\dagger},\hat{a}^{\dagger}]=0 \:.
  \end{equation}
However, on the other hand, the lowering (annihilation) and raising (creation) operators for fermions, $\hat{b}^{\dagger}$ and $\hat{b}$, respectively are defined as
  \begin{align} \label{a2.13}
 && \hat{b} &= \sqrt{\frac{1}{2 \hbar}} \left(\hat{\psi}_1 - i\hat{\psi}_2 \right), & 
  \hat{b}^{\dagger} &=\sqrt{\frac{1}{2 \hbar}} \left(\hat{\psi}_1 + i\hat{\psi}_2 \right) \:.&&
  \end{align}
Also, as will be shown in Section \ref{Sec2.4}, the fermionic operators $\psi_{\alpha}$ and $\psi_{\beta}$ satisfy the canonical quantum anticommutation condition $\{\psi_{\alpha},\psi_{\beta}\}= \hbar \delta_{\alpha \beta}$\footnote{The derivation of the fermionic anticommutators is a bit subtle and will be discussed in the next section.}. Therefore, it can be seen that the fermionic ladder operators $\hat{b}$ and $\hat{b}^{\dagger}$ satisfy the following anticommutation relations:
\begin{equation} \label{a2.18}
 \{ \hat{b},\hat{b}^{\dagger} \} =1, \quad  \quad  \{ \hat{b},\hat{b} \} = \{ \hat{b}^{\dagger},\hat{b}^{\dagger} \}=0 \:.
\end{equation}  
It is possible using Eq. (\ref{2.13}) to write the bosonic position and momentum variables in terms of the bosonic ladder operators as follows:
\begin{align} \label{c2.23}
&&&&&&&& \hat{a} + \hat{a}^\dagger &= \sqrt{\frac{2 \omega}{\hbar}} \hat{q} && &\Rightarrow & && \hat{q} = \sqrt{\frac{\hbar}{2 \omega}} (\hat{a}^\dagger+ \hat{a}), &&&&&&&& \nonumber \\
&&&&&&&& \hat{a} - \hat{a}^\dagger &= \sqrt{\frac{2}{\hbar \omega}} i \hat{p} && &\Rightarrow & && \hat{p} =i \sqrt{\frac{\hbar \omega}{2}} (\hat{a}^\dagger- \hat{a}) \:. &&&&&&&&
\end{align}
Similarly, using Eq. (\ref{a2.13}), we can write the fermionic variables $\psi_1$ and $\psi_2$ in terms of the fermionic ladder operators as follows:
\begin{align} \label{c2.24}
&&&&&&&& \hat{b}^\dagger + \hat{b} &= \sqrt{\frac{2}{\hbar}} \hat{\psi}_1 && &\Rightarrow & && \hat{\psi}_1 = \sqrt{\frac{\hbar}{2}} (\hat{b} + \hat{b}^\dagger), &&&&&&&& \nonumber \\
&&&&&&&& \hat{b}^\dagger - \hat{b} &= i \sqrt{\frac{2}{\hbar}}  \hat{\psi}_2 && &\Rightarrow & && \hat{\psi}_2 =i \sqrt{\frac{\hbar}{2}} (\hat{b} - \hat{b}^\dagger) \:. &&&&&&&&
\end{align}
Actually, we derived Eqs. (\ref{c2.23}, \ref{c2.24}) here; however, it will be used later.  At this stage, let us define the following two Hermitian operators, the bosonic number operator $\hat{\mathit{N}}_{B}$, and the fermionic number operator $\hat{\mathit{N}}_{F}$:
  \begin{align}\label{2.16}
  \mathit{\hat{N}}_B &\equiv \hat{a}^{\dagger} \hat{a} = \frac{1}    {2 \hbar \omega} \left(\omega^2 \hat{q}^2 - \hbar \omega + \hat{p}^2 \right),  \nonumber \\
   \mathit{\hat{N}}_F &\equiv \hat{b}^{\dagger} \hat{b} \:=
\frac{-i}{\hbar} \left(\hat{\psi}_1 \hat{\psi}_2 \right) \:.  
  \end{align} 
Using Eq. (\ref{2.16}), we can rewrite the Hamiltonian (\ref{2.9}) in the operator formalism 
  \begin{equation} \label{2.17}
  \mathit{\hat{H}} = \hbar \omega \left( \mathit{\hat{N}}_B + \frac{1}{2} \right) + \hbar \omega \left( \mathit{\hat{N}}_F - \frac{1}{2} \right) \:.
  \end{equation} 
This means that the energy eigenstates can be labeled with the eigenvalues of $\mathit{n}_B$ and $\mathit{n}_F$.

The action of the ladder operators, $\hat{a}$, $\hat{a}^{\dagger}$, $\hat{b}$ and $\hat{b}^{\dagger}$, upon an energy eigenstate $\vert n_B , n_F \rangle$ is listed as follows:
  \begin{align} \label{2.18}
& \resizebox{.46 \textwidth}{!}  {$\hat{a} \vert n_B, n_F \rangle= \sqrt{\mathit{n}_B} \vert    \mathit{n}_B -1, n_F \rangle, 
\quad   \hat{a}^\dagger \vert \mathit{n}_B, n_F \rangle= \sqrt{\mathit{n}_B+1} \vert \mathit{n}_B +1, n_F \rangle,$} \nonumber \\
 & \resizebox{.46 \textwidth}{!}  {$\hat{b} \vert n_B, \mathit{n}_F \rangle = \sqrt{\mathit{n}_F} \vert n_B, \mathit{n}_F -1 \rangle, 
\quad   \hat{b}^\dagger \vert n_B, \mathit{n}_F \rangle = \sqrt{\mathit{n}_F+1} \vert n_B, \mathit{n}_F +1 \rangle \:. $}
  \end{align}
It is straightforward from Eq. (\ref{2.18}) that the associated bosonic number operator $\mathit{\hat{N}}_B$ and fermionic number operator $\mathit{\hat{N}}_F$ obey
  \begin{equation} \label{2.19}
  \mathit{\hat{N}}_B \vert \mathit{n}_B, n_F \rangle = \mathit{n}_B \vert \mathit{n}_B, n_F \rangle,  \qquad   \mathit{\hat{N}}_F \vert n_B, \mathit{n}_F \rangle = \mathit{n}_F \vert n_B, \mathit{n}_F \rangle \:.
  \end{equation}
  
Under Eq. (\ref{2.19}), the energy spectrum of the Hamiltonian (\ref{2.17}) is
  \begin{equation} \label{2.20}
  \mathit{E}_n = \hbar \omega \left( \mathit{n}_B + \frac{1}{2} \right) + \hbar \omega \left( \mathit{n}_F - \frac{1}{2} \right) \:.
  \end{equation}
The form of Eq. (\ref{2.20}) implies that the Hamiltonian $\mathit{H}$ is symmetric under the interchange of $\hat{a}$ and $\hat{a}^{\dagger}$ and antisymmetric under the interchange of $\hat{b}$ and $\hat{b}^{\dagger}$.

The most important observation from Eq. (\ref{2.20}) is that $\mathit{E_n}$ remains invariant under a simultaneous annihilation of one boson ($\mathit{n}_B \rightarrow \mathit{n}_B -1$) and creation of one fermion ($\mathit{n}_F \rightarrow \mathit{n}_F + 1$) or vice versa. This is one of the simplest examples of a symmetry called "supersymmetry" (SUSY) and the corresponding energy spectrum reads
  \begin{equation}
  \mathit{E}_n= \hbar \omega \left( \mathit{n}_B+\mathit{n}_F \right) \:.
  \end{equation}
  
Moreover, we still have two more comments before ending this section:
\begin{itemize}
\item[(i)]
Consider that the fermionic vacuum state $\vert 0 \rangle$ is defined by
\begin{equation}
\hat{b} \vert 0 \rangle \equiv 0.
\end{equation}
Then, we define a fermionic one-particle state $\vert 1 \rangle$ by
\begin{equation}
\vert 1 \rangle \equiv \hat{b}^{\dagger} \vert 0 \rangle \:.
\end{equation}
It is now easy to use the anticommutation relations (\ref{a2.18}) to see that there are no other states, since operating on $\vert 1 \rangle$ with $\hat{b}$ or $\hat{b}^{\dagger}$ gives either the already known state $\vert 0 \rangle$, or nothing:
\begin{align}
& \resizebox{.42 \textwidth}{!} 
 {$\hat{b} \vert 1 \rangle = \hat{b} \hat{b}^{\dagger} \vert 0 \rangle = (1-\hat{b}^{\dagger} \hat{b}) \vert 0 \rangle = \vert 0 \rangle - \hat{b}^{\dagger} \hat{b} \vert 0 \rangle =  \vert 0 \rangle - 0 = \vert 0 \rangle,$} \nonumber \\
& \hat{b}^{\dagger} \vert 1 \rangle = \hat{b}^{\dagger} \hat{b}^{\dagger} \vert 0 \rangle = 0 \vert 0 \rangle =0 \:.
\end{align}
So, the subspace spanned by $\vert 0 \rangle$ and $\vert 1 \rangle$ is closed under the action of $\hat{b}$ and $\hat{b}^{\dagger}$, and, therefore, it is closed under the action of any product of these operators. We can show this directly using the two facts that the operator $\hat{b} \hat{b}^{\dagger}$ can be expressed as the linear combination $1 - \hat{b}^{\dagger} \hat{b}$ and $\hat{b}^2=\hat{b}^{{\dagger}^2}=0$. So, any product of three or more $\hat{b}$, $\hat{b}^{\dagger}$ can be shortened by using either $\hat{b}^2 = 0,$ or $\hat{b}^{{\dagger}^2} = 0$. For example $ \hat{b}^{\dagger} \hat{b} \hat{b}^{\dagger} = \hat{b}^{\dagger} - \hat{b} \hat{b}^{{\dagger}^2} = \hat{b}^{\dagger}$ and $\hat{b} \hat{b}^{\dagger} \hat{b} = \hat{b} - \hat{b}^{\dagger} \hat{b}^2 = \hat{b}$. So, it seems to be clear that all products of $\hat{b}$ and $\hat{b}^{\dagger}$ can always be reduced  to linear combinations of the following four operators $1$, $\hat{b}$, $\hat{b}^{\dagger}$, and $\hat{b}^{\dagger} \hat{b}$. Based on this argument, there are only two possible fermionic eigenstates and hence two possible eigenvalues of the fermionic number operator; they are $\mathit{n}_F=\{0,1\}$.
\item[(ii)] The ground eigenstate has a vanishing energy eigenvalue (when $\mathit{n}_B= \mathit{n}_F=0$). In this case, the ground eigenstate is not degenerate and we say that supersymmetry is unbroken. This zero-energy eigenvalue arises because of the cancellation between the bosonic and fermionic contributions to the supersymmetric ground-state energy since the ground-state energy for the bosonic and fermionic oscillators has the values  $\frac{1}{2} \hbar \omega$ and $-\frac{1}{2} \hbar \omega$, respectively.
\end{itemize}

\section{The Constrained Hamiltonian\\Formalism}

Before discussing the supersymmetry algebra in the next section, we have to discuss the constrained Hamiltonian formalism in this section. This formalism is needed to get the correct fermionic anticommutators due to the complication of constraints. Good discussions of the constrained Hamiltonian systems are presented in \cite{greenwood2006advanced, hanson1976constrained, marathe1983constrained, henneaux1992quantization, wipf1994hamilton, tavakoli2014lecture}. For instance, let us recall the Hamiltonian (\ref{2.6})
\begin{equation} \label{a2.31}
\mathit{H} = \frac{1}{2} \left( p^2 + \omega^2 q^2 \right) - i \omega \psi_1 \psi_2 \:.
\end{equation}
A constrained Hamiltonian is one in which the momenta and positions are related by some constraints. In general, $M$ constraints between the canonical variables can be written as
\begin{equation}
\phi_m(q,p,\psi,\pi)=0, \qquad   m=1, \dots , M.
\end{equation}
These are called primary constraints. Secondary constraints are additional constraints relating momenta and positions which can arise from the requirement that the primary constraints are time-independent; i.e., $\dot{\phi}_m = 0$. Fortunately, we do not have to deal with such constraints in this case. In principle, there can also be tertiary constraints arising from the equation of motion of the secondary constraints and so on.

The Hamiltonian (\ref{a2.31}) is an example of constrained Hamiltonians. In the previous section, we defined the fermionic momenta $\pi^\alpha$ in Eq. (\ref{2.8}) using the fermionic equation of motion followed from the usual Euler-Lagrange equations with the Lagrangian (\ref{2.5}) which is equivalent to this Hamiltonian. In fact, this definition gives us the relation between the fermionic momenta $\pi^\alpha$ and the canonical coordinates $\psi^\alpha$:
\begin{equation} \label{b2.34}
\pi^{\alpha} = \frac{\partial \mathit{L}}{\partial \dot{\psi}_{\alpha}}= - \frac{i}{2} \delta^{\alpha \beta} \psi_{\beta},  \qquad \alpha, \beta \in \{1,2 \} \:.
\end{equation}
In light of this equation, we have the primary constraint
\begin{equation} \label{b2.35}
\phi^\alpha = \pi^\alpha+ \frac{i}{2} \delta^{\alpha \beta} \psi_\beta = 0 \:.
\end{equation}
Such constraints mean that the way we write the Hamiltonian is ambiguous since we can exchange position and momentum variables which change the equations of motion one gets. For example, using Eq. (\ref{b2.34}) one can rewrite the Hamiltonian (\ref{a2.31}) in terms of the fermionic momenta as follows:  
\begin{equation} \label{a2.36}
\mathit{H} = \frac{1}{2} \left( p^2 + \omega^2 q^2 \right) +4i \omega \pi^1 \pi^2 \:.
\end{equation}
In the next subsection, the general properties of Poisson brackets and its relation with the Hamiltonian formalism in classical mechanics are briefly discussed. Later in the same subsection, the problem caused by the constrained Hamiltonians is described. Next, the Subsection \ref{SubSec.2.3.3} shows how to deal with this problem using the constraints.

\subsection{Poisson Brackets and First-Class Constraints } \label{SubSec2.3.1}

The Poisson bracket of two arbitrary functions $\mathit{F}(\hat{q},\hat{p}, \hat{\psi},\hat{\pi})$ and $\mathit{G}(\hat{q},\hat{p}, \hat{\psi},\hat{\pi})$ is defined as
\begin{equation}
\resizebox{.42 \textwidth}{!} {$
\{\mathit{F},\mathit{G}\}_P = \left( \dfrac{\partial \mathit{F}}{\partial q_i} \dfrac{\partial \mathit{G}}{\partial p^i} - \dfrac{\partial \mathit{F}}{\partial p^i}\dfrac{\partial \mathit{G}}{\partial q_i} \right)
 (-)^{\epsilon_\mathit{F}} \left( \dfrac{\partial \mathit{F}}{\partial \psi_\alpha} \dfrac{\partial \mathit{G}}{\partial \pi^\alpha} - \dfrac{\partial \mathit{F}}{\partial \pi^\alpha}\dfrac{\partial \mathit{G}}{\partial \psi_\alpha} \right) \:,$}
\end{equation}
where $\epsilon_\mathit{F}$ is $0$ if $\mathit{F}$ is Grassmann even and $1$ if $\mathit{F}$ is Grassmann odd\footnote{Grassmann-even variables refer to bosons, while Grassmann-odd variables refer to fermions.}. Actually, in our argument here, only the case of the odd Grassmann variables is considered. In general, the Poisson brackets have the following properties:
\begin{itemize}
\item[(i)] A constraint $\mathit{F}$ is said to be a first-class constraint if its Poisson  bracket with all the other constraints vanishes:
\begin{align} 
\{\phi^\mathit{A}, \mathit{F} \}_P =0 \:,  
\end{align} 
where $\mathit{A}$ runs over all the constraints. If a constraint is not a first class, it is a second class.
\item[(ii)] A Poisson bracket of two functions $\mathit{F}$ and $\mathit{G}$ is antisymmetric:
\begin{equation}
\{\mathit{F},\mathit{G}\}_P = - \{ \mathit{G}, \mathit{F} \}_P \:.
\end{equation}
\item[(iii)]  For any three functions $\mathit{A}$, $\mathit{B}$, and $\mathit{C}$, the Poisson bracket is linear in both entries:
\begin{equation}
\{ \mathit{A},\mathit{B}+\mathit{C}\}_P= \{\mathit{A},\mathit{B}\}_P + \{\mathit{A},\mathit{C}\}_P \:. 
\end{equation}
\item[(iv)] A Poisson bracket for any three functions satisfies the Leibniz rule:
\begin{equation}
\resizebox{.38 \textwidth}{!} {$
\{ \mathit{A},\mathit{B} \mathit{C}\}_P = \{ \mathit{A},\mathit{B}\}_P \mathit{C} + (-1)^{\epsilon_\mathit{A} \epsilon_\mathit{B}} \mathit{B} \{ \mathit{A},\mathit{C}\}_P, $}
\end{equation}
where $\epsilon_\mathit{F}$ is the Grassmann parity of $\mathit{F}$ which is $0$ if $\mathit{F}$ is Grassmann even and $1$ if $\mathit{F}$ is Grassmann odd.
\item[(v)] A Poisson bracket for any three functions obeys the Jacobi identity:
\begin{equation}
\{ \mathit{A}, \{\mathit{B},\mathit{C}\}_P \}_P + \{\mathit{B},\{\mathit{C},\mathit{A}\}_P\}_P + \{ \mathit{C}, \{ \mathit{A},\mathit{B}\}_P\}_P =0 \:.
\end{equation}
\item[(vi)] The time rate of change of any arbitrary function $\mathit{F}(q_i,p^i,\psi_\alpha,\pi^\alpha)$, which has no explicit time dependence, is given by its Poisson bracket with the Hamiltonian
\begin{equation}
\dot{\mathit{F}} = \{ \mathit{F}, \mathit{H} \}_P \:.
\end{equation}
\end{itemize}

In the Hamiltonian formalism of classical mechanics, the Hamilton equations of motion have equivalent expressions in terms of the Poisson bracket. Using the last property (vi) of the Poisson brackets, one can easily see that 
\begin{align} \label{bb2.44}
 \dot{q}_i &= \frac{\partial \mathit{H}}{\partial p^i} = \{ q_i, \mathit{H} \}_P, && &\dot{p}^i &=- \frac{\partial \mathit{H}}{\partial q_i} = \{ q^i, \mathit{H} \}_P,  \nonumber \\
\dot{\psi}_\alpha &= - \frac{\partial \mathit{H}}{\partial \pi^\alpha} = \{ \psi_\alpha, \mathit{H} \}_P, && &\dot{\pi}^\alpha &=- \frac{\partial \mathit{H}}{\partial \psi_\alpha} = \{ \pi^\alpha, \mathit{H} \}_P \:.
\end{align}
However, the problem with the constrained system is that the equations of motion derived using Eq. (\ref{bb2.44}) are not always consistent with those followed from the Lagrangian formalism. For example, if we consider the Hamiltonian (\ref{a2.31}), the equation of motion for $\psi_\alpha$ followed from the expected Poisson bracket equation has to be
\begin{equation} \label{b2.45}
\dot{\psi}_\alpha= \{ \psi_\alpha, \mathit{H} \}_P = - \frac{\partial \mathit{H}}{\partial \pi^\alpha} =0 \:.
\end{equation}
Unfortunately, this equation is inconsistent with the equation of motion followed from the Lagrangian formalism. Even worst, the equation of motion for $\pi^\alpha$ followed from the expected Poisson bracket equation is also inconsistent with the equation derived from the Lagrangian formalism. For example, using the Poisson bracket, we have
\begin{equation} \label{b2.46}
\dot{\pi}^1=\{ \pi^1,\mathit{H} \}_P = - \frac{\partial \mathit{H}}{\psi_1} = i \omega \psi_2 \:.
\end{equation}
But if we use Eq. (\ref{b2.34}) to find an expression to the equation of motion for $\dot{\psi}_1$ and then substitute into Eq. (\ref{b2.46}), that gives us 
\begin{equation}
\dot{\psi}_1 = 2i\dot{\pi}^1=-2\omega\psi_2 \:.
\end{equation}
This equation is different from the Lagrangian equations of motion and inconsistent with (\ref{b2.45}) (unless everything is trivial which is also not great). Moreover, using the Poisson bracket with the Hamiltonian (\ref{a2.36}), we can find the following set of equations of motion for the fermionic positions and momenta:
\begin{align}
\dot{\psi}_1 &= \{ \psi_1, \mathit{H} \}_P = - \frac{\partial \mathit{H}}{\partial \pi^1}= - 4i\omega \pi^2= - 2 \omega \psi_1, \nonumber \\
\dot{\psi}_2 &= \{ \psi_2, \mathit{H} \}_P = - \frac{\partial \mathit{H}}{\partial \pi^2}= - 4i\omega \pi^1= - 2 \omega \psi_2, \nonumber \\
\dot{\pi}^\alpha &= \{ \pi^\alpha, \mathit{H} \}_P = - \frac{\partial \mathit{H}}{\partial \psi_\alpha}= 0 \:.
\end{align}
The above equations are also slightly different from the Lagrangian equations of motion and inconsistent with the equations followed from the Hamiltonian (\ref{a2.31}).

In brief, there is an inconsistency between the equations of motion derived from the Hamiltonian formalism and the equations derived from the Lagrangian formalism. The reason for this inconsistency is that we have a constrained Hamiltonian. This problem is solved in the next subsection by imposing the constraints onto the Hamiltonian.

\subsection{The Constraints and the Equations of Motion} \label{SubSec.2.3.3}

In the Lagrangian formalism, one can impose the constraints from the beginning by introducing Lagrange multipliers which enforce the constraints. So, Hamilton's equations of motion, in the presence of constraints, can be derived from the extended action principle:
\begin{equation} \label{a2.39}
\mathit{S} = \int \left( \dot{q}_i p^i + \dot{\psi}_{\alpha} \pi^{\alpha} - \mathit{H}-\phi^A \lambda_A \right) dt \:,
\end{equation}
where $\phi^A$ are our constraints, $\lambda_A$ are Lagrange multipliers, and $\mathit{A}$ runs over the constraints. Using the extended action principle, we can write out the following two expressions of the extended Lagrangian and the extended Hamiltonian, respectively:
\begin{align}
\mathit{L} &= \dot{q}_i p^i + \dot{\psi}_{\alpha} \pi^{\alpha} - \mathit{H} - \phi^A \lambda_A \:, \\ \label{a2.41}
\mathit{H} &= \dot{q}_i p^i + \dot{\psi}_{\alpha} \pi^{\alpha} - \phi^A \lambda_A - \mathit{L} \:.
\end{align}
Indeed, if the Lagrangian is independent of one coordinate, say $q_i$, then, we call this coordinate an ignorable coordinate. However, we still need to solve the Lagrangian for this coordinate to find the corresponding equation of motion. Since the momentum corresponding to this coordinate may still enter the Lagrangian and affect the evolution of other coordinates, therefore, using the Euler-Lagrange equation, we have
\begin{equation}
\dfrac{d}{dt} \left( \dfrac{\partial \mathit{L}}{\partial \dot{q}_i} \right) - \dfrac{\partial \mathit{L}}{\partial q_i} = 0 \:.
\end{equation}
In general, we define the generalized momentum as $p^i = \partial \mathit{L} /\partial \dot{q}_i$. Now, if there is no explicit dependence of the Lagrangian $\mathit{L}$ on generalized coordinate $q_i$, then $\partial \mathit{L} / \partial q_i = 0$. Thus, Euler-Lagrange equation becomes $d/dt (\partial \mathit{L}/\partial \dot{q}_i) = 0 \: \Rightarrow \: d p^i/dt=0$. Hence, a momentum $p^i$ is conserved when the Lagrangian $\mathit{L}$ is independent of a coordinate $q_i$. This means that if the Lagrangian is independent of a certain coordinate $q^i$, it must be also independent of its corresponding momentum $p^i$. 

Furthermore, by extremizing the Hamiltonian (\ref{a2.41}), we obtain the following generalized Hamilton's equations of motion:
\begin{align}\label{b2.44}
 \dot{q}_i &=  \dfrac{\partial \mathit{H}}{\partial p^i } + \dfrac{\partial \phi^A}{\partial p^i} \lambda_A  = \{q_i  , \mathit{H} + \phi^A \lambda_A \}_P \:, \nonumber \\
 \dot{p}^i &= - \dfrac{\partial \mathit{H}}{\partial q_i} - \dfrac{\partial \phi^A}{\partial q_i} \lambda_A  = \{ p^i , \mathit{H} + \phi^A \lambda_A \}_P \:,  \nonumber \\
 \dot{\psi}_{\alpha} &= - \dfrac{\partial \mathit{H}}{\partial \pi^{\alpha}} - \dfrac{\partial \phi^A}{\partial \pi^{\alpha}} \lambda_A = \{\psi_{\alpha}  , \mathit{H} + \phi^A \lambda_A \}_P \:,  \nonumber \\
\dot{\pi}^{\alpha} &= - \dfrac{\partial \mathit{H}}{\partial  \psi_{\alpha}} - \dfrac{\partial \phi^A}{\partial \psi_{\alpha}} \lambda_A = \{\pi^{\alpha}  , \mathit{H} + \phi^A \lambda_A \}_P \:,  \nonumber \\
 \phi^A &= 0 \:.
\end{align}
Eqs. in (\ref{b2.44}) are derived in detail in \ref{AppendixA.1}. For example, using the Poisson brackets, we can show that Eqs. in (\ref{b2.44}) imply that for any arbitrary function $\mathit{F}= \mathit{F}(q_i,p^i,\psi_{\alpha},\pi^{\alpha})$, we can write $\dot{\mathit{F}}= \{ \mathit{F}, \mathit{H}+\phi^A \lambda_A \}_P$:
\begin{align} \label{b2.54}
\dot{\mathit{F}} &= \dfrac{\partial \mathit{F} }{\partial q_i} \dfrac{\partial q_i}{\partial t} + \dfrac{\partial \mathit{F} }{\partial p^i} \dfrac{\partial p^i}{\partial t} + \dfrac{\partial \mathit{F} }{\partial \psi_\alpha} \dfrac{\partial \psi_\alpha}{\partial t} + \dfrac{\partial \mathit{F} }{\partial \pi^\alpha} \dfrac{\partial \pi^\alpha}{\partial t} \nonumber \\
&= \dfrac{\partial \mathit{F} }{\partial q_i} \dot{q_i}+ \dfrac{\partial \mathit{F} }{\partial p^i} \dot{p^i} + \dfrac{\partial \mathit{F} }{\partial \psi_\alpha} \dot{\psi_\alpha} + \dfrac{\partial \mathrm{F} }{\partial \pi^\alpha} \dot{ \pi^\alpha} \nonumber \\ 
&= \dfrac{\partial \mathit{F} }{\partial q_i} \Big(  \dfrac{\partial \mathit{H}}{\partial p^i}+ \dfrac{\partial \phi^A}{\partial p^i} \lambda_A \Big) + \dfrac{\partial \mathit{F} }{\partial p^i} \Big(- \dfrac{\partial \mathit{H}}{\partial q^i} - \dfrac{\partial \phi^A}{\partial q^i} \lambda_A \Big) \nonumber \\
&{\quad}+ \dfrac{\partial \mathit{F} }{\partial \psi_\alpha} \Big(- \dfrac{\partial \mathit{H}}{\partial \pi^\alpha} - \dfrac{\partial \phi^A}{\partial \pi^\alpha} \lambda_A \Big) + \dfrac{\partial \mathit{F} }{\partial \pi^\alpha} \Big(- \dfrac{\partial \mathit{H}}{\partial \psi^\alpha} - \dfrac{\partial \phi^A}{\partial \psi_\alpha} \lambda_A \Big) \nonumber \\
&= \Big( \dfrac{\partial \mathrm{F} }{\partial q_i} \dfrac{\partial (\mathrm{H}+ \phi^A \lambda_A)}{\partial p^i}- \dfrac{\partial \mathrm{F} }{\partial p^i} \dfrac{\partial (\mathrm{H}+ \phi^A \lambda_A)}{\partial q^i} \Big) \nonumber \\
& \quad - \Big( \dfrac{\partial \mathrm{F} }{\partial \psi_\alpha} \dfrac{\partial (\mathrm{H}+ \phi^A \lambda_A)}{\partial \pi^\alpha} + \dfrac{\partial (\mathrm{H}+ \phi^A \lambda_A)}{\partial \psi^\alpha} \dfrac{\partial \mathrm{F} }{\partial \pi^\alpha}  \Big) \nonumber \\ 
& \equiv \{\mathrm{F}, \mathrm{H}+ \phi^A \lambda_A \}_P \:.
\end{align}
At this stage, we need to confirm that Eqs. in (\ref{b2.44}) give us the correct equations of motion using the constrained Hamiltonian (\ref{a2.31}). Since $\alpha, \beta \in \{1,2\}$ in the primary constraint (\ref{b2.35}), then, we have the two constraints
\begin{align} \label{aa2.57}
&&&&&&\phi^1 &= \pi^1 + \frac{i}{2} \psi_1, && \phi^2 = \pi^2 + \frac{i}{2} \psi_2. &&&&&&
\end{align}
By using Eqs. in (\ref{b2.44}) and taking into count the Hamiltonian (\ref{a2.36}), and the constraints (\ref{aa2.57}), we find that
\begin{align} \label{aa2.59}
\dot{\psi}_1 &= - \dfrac{\partial \mathit{H}}{\partial \pi^1} - \dfrac{\partial \phi^1}{\pi^1} \lambda_1 = -  \lambda_1, \nonumber \\
\dot{\psi}_2 &= - \dfrac{\partial \mathit{H}}{\partial \pi^2} - \dfrac{\partial \phi^2}{\pi^2} \lambda_2 = - \lambda_2 \:. 
\end{align}
On the other hand, using Eqs. in (\ref{b2.44}) and considering the Hamiltonian (\ref{a2.31}) and the constraints (\ref{aa2.57}), we find that
\begin{align} \label{aa2.60}
\dot{\pi}^1 &= - \dfrac{\partial \mathit{H}}{\partial \psi_1} - \dfrac{\partial \phi^1}{\psi_1} \lambda_1 = i \omega \psi_2 - \frac{i}{2} \lambda_1, \nonumber \\
\dot{\pi}^2 &= - \dfrac{\partial \mathit{H}}{\partial \psi_2} - \dfrac{\partial \phi^2}{\psi_2} \lambda_2 = -i \omega \psi_1 - \frac{i}{2} \lambda_2 \:.
\end{align}
Furthermore, using Eq. (\ref{b2.34}), we get
\begin{align} \label{aa2.62}
&&&&&&&&\pi^1 &= - \frac{i}{2} \psi_1 && \Rightarrow && \dot{\pi^1} = - \frac{i}{2} \dot{\psi}_1, &&&&&&&&  \nonumber \\
&&&&&&&&\pi^2 &= - \frac{i}{2} \psi_2 && \Rightarrow && \dot{\pi^2} = - \frac{i}{2} \dot{\psi}_2 \:. &&&&&&&&
\end{align}
Eqs. (\ref{aa2.59}, \ref{aa2.60}, and \ref{aa2.62}) imply the following set of equations of motion for the fermionic coordinates $\psi_\alpha$:
\begin{align} \label{aa2.63}
&&&&&& \dot{\psi}_1 &= - \omega \psi_2, && \dot{\psi}_2 =  \omega \psi_1 \:. &&&&&&
\end{align}
Fortunately, these equations are the same equations of motion followed from the Lagrangian formalism.

For now, let us use the Poisson bracket, the Hamiltonian (\ref{a2.31}), and the constraints (\ref{aa2.57}), to check that
\begin{align} \label{aa2.64}
 \dot{\phi}^1  & = \{ \phi^1, \mathit{H}+\phi^1 \lambda_1   \}_P \nonumber \\
& =  \left( \dfrac{\partial \phi^1}{\partial q_i} \dfrac{\partial (\mathit{H}+\phi^1 \lambda_1) }{\partial p^i} - \dfrac{\partial \phi^1}{\partial p^i} \dfrac{\partial (\mathit{H}+\phi^1 \lambda_1)}{\partial q_i} \right) \nonumber \\
& =  -  \left( \dfrac{\partial \phi^1}{\partial \psi_\alpha} \dfrac{\partial (\mathrm{H}+\phi^1 \lambda_1) }{\partial \pi^\alpha} + \dfrac{\partial \phi^1}{\partial \pi^\alpha} \dfrac{\partial (\mathit{H}+\phi^1 \lambda_1)}{\partial \psi_\alpha} \right) \nonumber \\
& =  0+0 - \Big(\frac{i}{2}\Big) \left(\lambda_1\right) - \left(1\right) \Big(-i\omega \psi_2+\frac{i}{2} \lambda_1 \Big) \nonumber \\
&= i \left(- \lambda_1+\omega \psi_2 \right)=0 \:.
\end{align}
The reason that the last step is zero comes from Eqs. (\ref{aa2.59}, \ref{aa2.63}) since we have $\lambda_1= - \dot{\psi}_1 = \omega \psi_2$. Similarly, $\dot{\phi}^2$ can be calculated using the Poisson bracket as follows:
\begin{align} \label{aa2.65}
 \dot{\phi}^2 &= \{ \phi^2, \mathit{H}+\phi^2 \lambda_2 \}_P \nonumber \\
& = \left( \dfrac{\partial \phi^2}{\partial q_i} \dfrac{\partial (\mathit{H}+\phi^2 \lambda_2) }{\partial p^i} - \dfrac{\partial \phi^2}{\partial p^i} \dfrac{\partial (\mathit{H}+\phi^2 \lambda_2)}{\partial q_i} \right) \nonumber \\
& = - \left( \dfrac{\partial \phi^2}{\partial \psi_\alpha} \dfrac{\partial (\mathit{H}+\phi^2 \lambda_2) }{\partial \pi^\alpha} + \dfrac{\partial \phi^2}{\partial \pi^\alpha} \dfrac{\partial (\mathit{H}+\phi^2 \lambda_2)}{\partial \psi_\alpha} \right) \nonumber \\
&=  0 + 0 - \Big( \frac{i}{2} \Big) \left( \lambda_2 \right) - \left(2\right) \Big( i\omega \psi_1+\frac{i}{2} \lambda_2 \Big)  \nonumber \\
&= i \left(- \lambda_2-\omega \psi_1 \right)=0 \:.
\end{align}
Also, the last step is zero because from Eqs. (\ref{aa2.59}, \ref{aa2.63}), we have $\lambda_2= - \dot{\psi}_2 = - \omega \psi_1$. Subsequently, Eqs. (\ref{aa2.64}, \ref{aa2.65}) are zero if we impose the equations of motion (\ref{aa2.63}). In other words, we can say that the constraints are consistent with the equations of motion.

\subsection{Dirac Brackets and Second Class Constraints }

A Constraint $\mathit{F}$ is said to be a second class constraint if it has nonzero Poisson brackets, and, therefore, it requires special treatment. Given a set of second class constraints, we can define a matrix
\begin{equation}
\{ \phi^{\mathit{A}},\phi^{\mathit{B}} \}_P = \mathit{C}^{AB} \:,
\end{equation}
where we define the inverse of $\mathit{C}^{\mathit{A} \mathit{B}},$ as $\mathit{C}_{\mathit{A} \mathit{B}}$, such that \label{b2.47}
\begin{equation}
\mathit{C}^{\mathit{A} \mathit{B}} \mathit{C}_{\mathit{B} \mathit{C}} = \delta^{\mathit{A}}_{\mathit{C}} \:.
\end{equation}
The Dirac bracket provides a modification to the Poisson brackets to ensure that the second class constants vanish:
\begin{equation}\label{a2.47}
\{ \phi^A, \mathit{F} \}_D =0 \:.
\end{equation}
The Dirac bracket of two arbitrary functions $F(q,p)$ and $G(q,p)$ is defined as
\begin{equation} \label{a2.48}
\{ F , G \}_D = \{ F,G \}_P - \{ F, \phi^A\}_P \mathit{C}_{AB} \{ \phi^B , G\}_P \:. 
\end{equation}

Dirac bracket has the same properties of Poisson bracket, which we have listed in Subsection \ref{SubSec2.3.1}. In addition, we have to notice the following.
\begin{itemize} 
\item[(i)] If $\dot{\phi}^A=0$, then, the Dirac bracket of any constraint with the extended Hamiltonian is equivalent to its Poisson bracket:
\begin{equation}
\{ \mathit{F}, \mathit{H} + \phi^A \lambda_A \}_D = \{ \mathit{F}, \mathit{H} + \phi^A \lambda_A \}_P \:.
\end{equation}
\item[(ii)] In some cases, we can return to the original Hamiltonian (\ref{a2.31}) forget about the constraints and the momenta $\pi^{\beta}$ at the cost of replacing the Poisson brackets with Dirac brackets. For example, if we consider the case $\dot{\phi}^A =0$, then, we have 
\begin{align}
& \resizebox{0.43 \textwidth}{!} {$
\dot{\mathit{F}}  = \{ \mathit{F}, \mathit{H}+\phi^A \lambda_A \}_D  = \{ \mathit{F}, \mathit{H} \}_D + \{ \mathit{F},\phi^A \}_D \lambda_A = \{ \mathit{F}, \mathit{H} \}_D \:,$}
\end{align} 
where we used Eq. (\ref{a2.47}); $\{\phi^A, \mathit{F} \}_D = 0$.
\item[(iii)] If $\{\phi^A,p\}_P=0$, then, using Eq. (\ref{a2.48}), we can find that
\begin{align}
\{ q,p\}_D & = \{ q,p\}_P - \{ q, \phi^A \}_P \mathit{C}_{AB} \{ \phi^B, p\}  \nonumber \\
& =  \{ q,p\}_P = \dfrac{\partial q}{\partial q} \dfrac{\partial p}{\partial p} - \dfrac{\partial q}{\partial p} \dfrac{\partial p}{\partial q} = 1 \:.
\end{align}
\end{itemize}

We are now in a position to calculate $\{ \psi_{\alpha} , \psi_{\beta} \}_D$ for the Hamiltonian (\ref{a2.31}) with the constraints (\ref{b2.35}). That $\{ \psi_{\alpha} , \psi_{\beta} \}_D$ involves a primary constraint $\phi^\beta$ followed from the following relation which is obtained from the Lagrangian formalism:
\begin{equation} \label{b2.65}
\phi^\beta = \pi^\beta + \frac{i}{2} \delta^{\alpha \beta} \psi_\alpha \:.
\end{equation}
Using the definition of Dirac bracket, we  get 
\begin{equation} \label{b2.64}
\{ \psi_{\alpha} , \psi_{\beta} \}_D = \{ \psi_{\alpha} , \psi_{\beta} \}_P - \{ \psi_{\alpha} , \phi^{\alpha} \}_P \mathit{C}_{\alpha \beta} \{ \phi^{\beta}, \psi_{\beta} \}_P \:.
\end{equation}
Then, since the variables $\psi_{\alpha}$ and $\psi_{\beta}$ are independent, their Poisson bracket vanishes; i.e.,
\begin{equation} \label{b2.66}
\{ \psi_{\alpha} , \psi_{\beta} \}_P = \{ \psi_{\beta} , \psi^{\alpha} \}_P = 0 \:.
\end{equation}
Furthermore, using Eq. (\ref{b2.54}), we find that
\begin{align} \label{b2.67}
\{ \psi_{\alpha}, \phi^{\alpha} \}_P &=  \left( \frac{\partial \psi_{\alpha}}{\partial q_i } \frac{\partial \phi^{\alpha}}{\partial p^i} - \frac{\partial \psi_{\alpha}}{\partial p^i} \frac{\partial \phi^{\alpha}}{\partial q_i} \right) - \left( \frac{\partial \psi_{\alpha}}{\partial \psi_\alpha } \frac{\partial \phi^{\alpha}}{\partial \pi^\alpha} + \frac{\partial \psi_{\alpha}}{\partial \pi^\alpha} \frac{\partial \phi^{\alpha}}{\partial \psi_\alpha} \right)   \nonumber \\
&=-1 \:.
\end{align}
Similarly, using Eq. (\ref{b2.65}), we find that
\begin{align} \label{b2.68}
\{ \psi_{\beta}, \phi^{\beta} \}_P &= \left( \frac{\partial \psi_{\beta}}{\partial q_i } \frac{\partial \phi^{\beta}}{\partial p^i} - \frac{\partial \psi_{\beta}}{\partial p^i} \frac{\partial \phi^{\beta}}{\partial q_i} \right) 
- \left( \frac{\partial \psi_{\beta}}{\partial \psi_\beta } \frac{\partial \phi^{\beta}}{\partial \pi^\beta} + \frac{\partial \psi_{\beta}}{\partial \pi^\beta} \frac{\partial \phi^{\beta}}{\partial \psi_\beta} \right)   \nonumber \\
&=-1 \:.
\end{align}
In addition, using the constraints, definitions (\ref{b2.54}, \ref{b2.65}), we can define a matrix
\begin{align}
\mathit{C}^{ \alpha \beta} &= \{ \phi^\alpha, \phi^\beta\}_P \nonumber \\
&= \resizebox{.42 \textwidth}{!} {$\left( \frac{\partial \phi^{\alpha}}{\partial q_i } \frac{\partial \phi^{\beta}}{\partial p^i} - \frac{\partial \phi^{\alpha}}{\partial p^i} \frac{\partial \phi^{\beta}}{\partial q_i} \right)
- \left( \frac{\partial \phi^{\alpha}}{\partial \psi_\alpha } \frac{\partial \phi^{\beta}}{\partial \pi^\alpha} + \frac{\partial \phi^{\alpha}}{\partial \pi^\alpha} \frac{\partial \phi^{\beta}}{\partial \psi_\alpha} \right) $} \nonumber \\
& \quad \:\:\:  \resizebox{.21 \textwidth}{!}  {$- \left( \frac{\partial \phi^{\alpha}}{\partial \psi_\beta } \frac{\partial \phi^{\beta}}{\partial \pi^\beta} + \frac{\partial \phi^{\alpha}}{\partial \pi^\beta} \frac{\partial \phi^{\beta}}{\partial \psi_\beta} \right)$} \nonumber \\
&= -i \delta^{\alpha \beta} \:.
\end{align}
Moreover, from the definition (\ref{b2.47}), we have
\begin{equation} \label{b2.70}
\mathit{C}_{\alpha \beta} = -\mathit{C}^{\alpha \beta} = i \delta^{\alpha \beta} \:.
\end{equation} 
As a result of substituting Eqs. (\ref{b2.66}, \ref{b2.67}, and \ref{b2.68}, \ref{b2.70}) into Eq. (\ref{b2.64}), the Dirac bracket of the two variables $\psi_{\alpha}$ and $\psi_{\beta}$ gives us
\begin{equation}
\{ \psi_{\alpha} , \psi_{\beta} \}_D = i \delta^{\alpha \beta} \:.
\end{equation}

\subsection{Dirac Bracket and the Equations of Motion} 

In this subsection, we show that the Dirac brackets give the correct equations of motion for the Hamiltonian (\ref{b2.34}) using the expression
\begin{equation}
\dot{\psi}_{\alpha} = \{ \psi_{\alpha} , \mathit{H} \}_D \:,
\end{equation}
We check that
\begin{align}
\dot{\psi}_1 & = \{ \psi_1, \mathit{H} \}_D = \{ \psi_1, - i\omega \psi_1 \psi_2 \}_D = -i\omega \{ \psi_1,\psi_1 \psi_2 \}_D \nonumber \\
&= -i\omega \left( \{ \psi_1,\psi_1 \}_D \psi_2 + \psi_1 \{ \psi_1,\psi_2 \}_D \right) = -i\omega \left( -i \delta_{\alpha \beta} \psi_2 + 0 \right) \nonumber \\
&= - \omega \delta_{\alpha \beta} \psi_2 = - \omega \psi_2 \:,
\end{align}
and similarly
\begin{align}
\dot{\psi}_2 &= \{ \psi_2, \mathit{H} \}_D = \{ \psi_2, - i\omega \psi_1 \psi_2 \}_D = -i\omega \{ \psi_2,\psi_1 \psi_2 \}_D \nonumber \\ 
&= \resizebox{.42 \textwidth}{!} {$ -i\omega \left( \{ \psi_2,\psi_1 \}_D \psi_2 + \psi_1 \{ \psi_2,\psi_2 \}_D \right) = -i\omega \left( 0 + \psi_1 (-i \delta_{\alpha \beta}) \right)$} \nonumber \\
&= - \omega \psi_1 \delta_{\alpha \beta}  =  \omega \delta_{\alpha \beta} \psi_1  = \omega \psi_1 \:.  
\end{align}
These equations are the same equations of motion which we have obtained in Subsection \ref{SubSec.2.3.3}  using the constrained Hamilton's equations of motion \ref{b2.44}.
 
\subsection{The Dirac Bracket Superalgebra} \label{SubSec2.3.5}

In this subsection, we investigate the supersymmetry of our Hamiltonian (\ref{a2.31}). For this purpose, we need to introduce some operators $\mathit{Q}_i$ known as supercharges, which basically can be obtained from N\"{o}ether's theorem arising from a symmetry of the Lagrangian which exchanges bosons and fermions. In the next section, we will look at the action of the supercharge operators. For instance, the supercharge operators in the case of the system under discussion here can be written as follows:
\begin{align}\label{q2.71}
\mathit{Q}_\alpha  = p \psi_\alpha + \omega q \epsilon_{\alpha \beta} \delta^{\beta \gamma} \psi_\gamma \quad \Rightarrow \quad 
\begin{split} 
\mathit{Q}_1  &= p \psi_1 + \omega q \psi_2  \\
\mathit{Q}_2  &= p \psi_2 - \omega q \psi_1 \:,
\end{split}
\end{align}
where again we suppose that $\{\alpha,\beta\}$ can only take the values $\{1,2\}$, and $\epsilon_{\alpha \beta}$ is the two index Levi-Civita symbol defined as  
\begin{equation}
\epsilon_{\alpha \beta }= \begin{cases} \; \; +1 \quad \texttt{if} \: (\alpha ,\beta) \: \texttt{is} \: (1,2), \\ \; \; -1 \quad \texttt{if} \: (\alpha ,\beta) \:\texttt{is} \: (2,1), \\ \; \; 0 \quad \: \: \: \, \texttt{if} \: \alpha=\beta \:. \end{cases}
\end{equation}
The system to be supersymmetric, the supercharges (\ref{q2.71}) together with the Hamiltonian (\ref{b2.34}) and the constraints (\ref{a2.31}) must satisfy the classical Dirac bracket superalgebra which defined by
\begin{align} \label{aa2.73}
\{ \mathit{Q}_{\alpha}, \mathit{Q}_{\beta} \}_D & = - 2 i \delta_{\alpha \beta} \mathit{H}, \nonumber \\
\{ \mathit{Q}_{\alpha}, \mathit{H} \}_D &=0 \:.
\end{align}  

Now, we need to check that our system satisfies Eq. (\ref{aa2.73}). To do that, suppose that $\{\alpha , \beta\}$ have the values $\{1,2\}$. Then, using the definition of the Poisson and Dirac brackets, we find that,(see \ref{AppendixA.2} for more details)
\begin{align}\label{aa2.74}
\{ \mathit{Q}_1, \mathit{Q}_1 \}_D &= -i \left( p^2+\omega^2 q^2 \right), && \{ \mathit{Q}_1, \mathit{Q}_2 \}_D = \omega \left( \psi_1^2+\psi_2^2 \right), \nonumber \\ 
\{ \mathit{Q}_2, \mathit{Q}_1 \}_D &= - \omega \left( \psi_1^2+\psi_2^2 \right), && \{ \mathit{Q}_2, \mathit{Q}_2 \}_D = -i \left( p^2+\omega^2 q^2 \right) \:.
\end{align}
If we make substitution using Eq. (\ref{aa2.74}), we find 
\begin{align} \label{aa2.75}
\resizebox{.08 \textwidth}{!}  {$\{Q_\alpha, Q_\beta \}_D$} & = \resizebox{.38 \textwidth}{!} {$ \{Q_1, Q_1 \}_D + \{Q_1, Q_2 \}_D + \{Q_2, Q_1 \}_D + \{Q_2, Q_2 \}_D $} \nonumber \\
& = \resizebox{.38 \textwidth}{!} {$ - i ( p^2 + \omega^2 q^2) + \omega (\psi_1^2 + \psi_2^2)  - \omega (\psi_1^2 + \psi_2^2) - i ( p^2 + \omega^2 q^2)$} \nonumber \\
&= - 2i ( p^2 + \omega^2 q^2) \equiv -2 i \delta_{\alpha \beta} \mathit{H} \:.
\end{align}
Furthermore, using the following relations (see \ref{AppendixA.2})
\begin{align}\label{aa2.76}
\{ \mathit{Q}_1, \mathit{H} \}_P & = \omega p \psi_2 - \omega^2 q \psi_1, && \{ \mathit{Q}_2 \:, \mathit{H} \}_D = - \omega p \psi_2 + \omega^2 q \psi_1 \:, \nonumber \\ 
\{ \phi_1, \mathit{H} \}_P &= i \omega \psi_2, && \{ \phi_2, \mathit{H} \}_P = i \omega \psi_2 \:,
\end{align}
we find 
\begin{align} \label{aa2.77}
\resizebox{.07 \textwidth}{!} {$ \{Q_\alpha, \mathit{H}\}_D $}& = \resizebox{.16 \textwidth}{!}  {$ \{Q_1, \mathit{H}\}_D + \{Q_2, \mathit{H}\}_D $} \nonumber \\
& =\resizebox{.25 \textwidth}{!} {$ \{Q_1, \mathit{H}\}_P - \{ Q_1, \phi^A \}_P C_{AB} \{ \phi^B, \mathit{H} \}_P $} \nonumber \\
& \quad \resizebox{.25 \textwidth}{!}  {$+ \{Q_2, \mathit{H}\}_P - \{ Q_2, \phi^A \}_P C_{AB} \{ \phi^B, \mathit{H} \}_P $}  \nonumber \\
& =\resizebox{.38 \textwidth}{!}  {$ \{Q_1, \mathit{H}\}_P - \{ Q_1, \phi^1 \}_P C_{11} \{ \phi^1, \mathit{H} \}_P - \{ Q_1, \phi^2 \}_P C_{22} \{ \phi^2, \mathit{H} \}_P $} \nonumber \\
& \quad \resizebox{.39 \textwidth}{!} {$ + \{Q_2, \mathit{H}\}_P - \{ Q_2, \phi^1 \}_P C_{11} \{ \phi^1, \mathit{H} \}_P - \{ Q_2, \phi^2 \}_P C_{22} \{ \phi^2, \mathit{H} \}_P $} \nonumber \\
& =   \resizebox{.23 \textwidth}{!}  {$\omega P \psi_2 - \omega^2 q \psi_1 - \omega P \psi_2 + \omega^2 q \psi_1 $} \nonumber \\
& \quad \resizebox{.23 \textwidth}{!} {$ - \omega P \psi_1 - \omega^2 q \psi_2 + \omega^2 q \psi_2 + \omega P \psi_1$}  \nonumber \\
& = 0 \:.
\end{align}
It is clear from Eqs. (\ref{aa2.75}, \ref{aa2.77}) that the supercharges (\ref{q2.71}) together with the Hamiltonian (\ref{b2.34}) and the constraints (\ref{b2.35}) satisfy the two conditions (\ref{aa2.73}) of the classical Dirac bracket superalgebra.

\section{The Supersymmetry Algebra} \label{Sec2.4}

The material in this section is elaborated in detail in \cite{kugo1983supersymmetry,drees1996introduction, bilal2001introduction,combescure2004n, bougie2012supersymmetric,labelle2010supersymmetry}. The supersymmetry algebra encodes a symmetry describing a relation between bosons and fermions. In general, the supersymmetry is constructed by introducing supersymmetric transformations which are generated by the supercharge $\mathit{Q}_i$ operators, where the role of the supercharges $\mathit{Q}_i$ is to convert a fermionic degree of freedom into a bosonic degree of freedom and vice versa; i.e.,
\begin{equation}
  \mathit{Q} \vert \texttt{fermionic} \rangle = \vert \texttt{bosonic} \rangle , \qquad \mathit{Q} \vert \texttt{bosonic} \rangle= \vert \texttt{fermionic} \rangle \:.
\end{equation}
   
So far, we have restricted ourselves to study classical systems. In Subsection \ref{SubSec2.3.5}, we have investigated the supersymmetry of the classical harmonic oscillator. Now, we are going to quantize the theory to study the supersymmetry of quantum systems. In a quantum mechanical supersymmetric system, the supercharges $\mathit{Q}_i$ together with the Hamiltonian $\mathit{H}$ form a so-called superalgebra. The recipe for quantizing the Hamiltonian in a situation where we have second class constraints is to replace the Dirac brackets with either commutator brackets for bosonic variables or anticommutator brackets for fermionic variables multiplied with the factor $i \hbar$, so we have
\begin{align} \label{d2.90}
&&&&&&&& \{ q,p\}_D &= 1 & \Rightarrow & \qquad &[ \hat{q}, \hat{p}] &= i \hbar, &&&&&&&& \nonumber \\
&&&&&&&& \{\psi_{\alpha}, \psi_{\beta} \}_D &= -i \delta_{\alpha \beta} & \Rightarrow & \qquad &\{ \hat{\psi}_{\alpha}, \hat{\psi}_{\beta} \} &= \hbar \delta_{\alpha \beta} \:. &&&&&&&&
\end{align}
Taking this into account, we can see that the superalgebra for N-dimensional quantum system is characterized by
\begin{align} \label{2.24}
 &&&&&&       [ \hat{\mathit{Q}}_i, \hat{\mathit{H}}] &= 0, & i &= 1 \cdots \mathit{N}, &&&&&& \nonumber \\  &&&&&&      
\{ \hat{\mathit{Q}}_i, \hat{\mathit{Q}}_j \} &= \hbar \delta_{ij} \hat{\mathit{H}}, & i,j &= 1     \dots \mathit{N} \:. &&&&&&
\end{align}
This will be elaborated further in the next section using the example of the supersymmetrical harmonic oscillator. One more notation before ending this section is that the supercharges $\mathit{Q}_i$ are Hermitian, i.e. $\mathit{Q_i^{\dagger}}=\mathit{Q_i}$, and this implies that
  \begin{equation}
  \{ \mathit{\hat{Q}}_i,\mathit{\hat{Q}}_j \}=  \{ \mathit{\hat{Q}}_i^{\dagger},\mathit{\hat{Q}}_j^{\dagger} \} \:.
  \end{equation}

\section{Supersymmetric Harmonic Oscillator} \label{Sec2.5}

Now, we turn back to the quantum harmonic oscillator problem as a simple example to show the supersymmetric property of a quantum mechanical system. Using Eq. (\ref{2.16}), the harmonic oscillator Hamiltonian (\ref{2.17}) can be written in terms of the bosonic and fermionic laddering operators as follows:
\begin{equation} \label{c2.93}
\mathit{\hat{H}} = \hbar \omega ( \hat{a}^\dagger \hat{a} + \hat{b}^\dagger \hat{b}) \:.
\end{equation}
Moreover, the supercharge operators $\mathit{\hat{Q}}_1$ and $\mathit{\hat{Q}}_2$ can also be written in terms of the bosonic and fermionic laddering operators as follows:
\begin{align} \label{c2.94}
\mathit{\hat{Q}}_1 &= p \psi_1 + \omega q \psi_2 \nonumber \\
&= \left( i \sqrt{\frac{\hbar \omega}{2}} (\hat{a}^\dagger - \hat{a}) \right) \left( \sqrt{\frac{\hbar}{2}} (\hat{b} + \hat{b}^\dagger) \right)  \nonumber \\
& \quad + \omega \left( \sqrt{\frac{\hbar}{2 \omega}} (\hat{a})^\dagger + \hat{a} ) \right) \left( i \sqrt{\frac{\hbar}{2}} (\hat{b}-\hat{b}^\dagger) \right) \nonumber \\
&= i \hbar \sqrt{\omega} (\hat{a}^\dagger \hat{b}- \hat{a} \hat{b}^\dagger ) \:,
\end{align}
and similarly,
\begin{align} \label{c2.95}
\mathit{\hat{Q}}_2 &= p \psi_2 - \omega q \psi_1 \nonumber \\
&= \left( i \sqrt{\frac{\hbar \omega}{2}} (\hat{a}^\dagger - \hat{a}) \right) \left( i \sqrt{\frac{\hbar}{2}} (\hat{b}-\hat{b}^\dagger) \right) \nonumber \\
& \quad- \omega \left( \sqrt{\frac{\hbar}{2 \omega}} (\hat{a})^\dagger + \hat{a} ) \right) \left( \sqrt{\frac{\hbar}{2}} (\hat{b} + \hat{b}^\dagger) \right) \nonumber \\
&= - \hbar \sqrt{\omega} (\hat{a}^\dagger \hat{b}+ \hat{a} \hat{b}^\dagger) \:.
\end{align}

Using Eqs. (\ref{c2.94}, \ref{c2.95}), one may define non-Hermitian operators $\mathit{\hat{Q}}$ and $\mathit{\hat{Q}}^\dagger$ as
\begin{align} \label{c2.96}
\mathit{\hat{Q}} &= \frac{1}{2} ( \mathit{\hat{Q}}_1 -i \mathit{\hat{Q}}_2 ) \nonumber \\
&=  \frac{1}{2} \left( i \hbar \sqrt{\omega} (\hat{a}^\dagger \hat{b}- \hat{a} \hat{b}^\dagger ) + i\hbar \sqrt{\omega} (\hat{a}^\dagger \hat{b}+ \hat{a} \hat{b}^\dagger ) \right) \nonumber \\
&= i \hbar \sqrt{\omega} (\hat{a}^\dagger \hat{b}) \:,
\end{align} 
and similarly,
\begin{align} \label{c2.97}
\mathit{\hat{Q}}^\dagger &= \frac{1}{2} ( \mathit{\hat{Q}}_1 + i \mathit{\hat{Q}}_2 ) \nonumber \\
&=  \frac{1}{2} \left( i \hbar \sqrt{\omega} (\hat{a}^\dagger \hat{b}- \hat{a} \hat{b}^\dagger ) - i\hbar \sqrt{\omega} (\hat{a}^\dagger \hat{b}+ \hat{a} \hat{b}^\dagger ) \right) \nonumber \\
&= - i \hbar \sqrt{\omega} (\hat{a} \hat{b}^\dagger) \:.
\end{align} 
We can directly use Eq. (\ref{c2.93}, \ref{c2.96}, \ref{c2.97}) to check that the supercharge operators $\mathit{\hat{Q}}$ and $\mathit{\hat{Q}}^\dagger$ with the Hamiltonian $\mathit{H}$ satisfy the quantum superalgebra:
\begin{align} \label{c2.98}
&&&&&&&&&&&&&&&& \{ \mathit{\hat{Q}}, \mathit{\hat{Q}} \} & =& \{ \mathit{\hat{Q}}^\dagger , \mathit{\hat{Q}}^\dagger \} &= 0, &&&&&&&&&&&&&&&& \nonumber \\
&&&&&&&&&&&&&&&& \{ \mathit{\hat{Q}}, \mathit{\hat{Q}}^\dagger \} &=& \{ \mathit{\hat{Q}}^\dagger , \mathit{\hat{Q}} \} &= \mathit{H},&&&&&&&&&&&&&&&& \nonumber \\
&&&&&&&&&&&&&&&& [ \mathit{H}, \mathit{\hat{Q}} ] &= &[ \mathit{H}, \mathit{\hat{Q}}^\dagger ] &= 0 \:. &&&&&&&&&&&&&&&&
\end{align} 

Now, it is easy using Eqs. (\ref{c2.96}, \ref{c2.97}) to see the action of the operators $\mathit{\hat{Q}}, \mathit{\hat{Q}}^{\dagger}$ on the energy eigenstates:
\begin{align}
\mathit{\hat{Q}} \vert \mathit{n}_{B}, \mathit{n}_{F} \rangle & \sim  \vert \mathit{n}_{B}-1, \mathit{n}_{F}+1 \rangle, \nonumber  \\
\mathit{\hat{Q}}^{\dagger} \vert \mathit{n}_{B}, \mathit{n}_{F} \rangle & \sim  \vert \mathit{n}_{B}+1, \mathit{n}_{F}-1 \rangle \:.
\end{align}
Remember that there are only two possible fermionic states $\mathit{n}_{F}=\{0,1\}$, so the effect of the operators $\mathit{\hat{Q}}$ and $\mathit{\hat{Q}}^{\dagger}$ can be written as follows:
\begin{align}
\mathit{\hat{Q}} \vert \mathit{n}_{B}, 0 \rangle & \sim \vert \mathit{n}_{B}-1, 1 \rangle, \nonumber  \\
\mathit{\hat{Q}}^{\dagger} \vert \mathit{n}_{B}, 1 \rangle & \sim  \vert \mathit{n}_{B}+1, 0 \rangle \:.
\end{align}
Also, from Eq. (\ref{c2.98}),  the operators $\mathit{\hat{Q}}$ and $\mathit{\hat{Q}}^\dagger$ commute with the Hamiltonian $\mathit{H}$; for that reason, we can see that
\begin{equation}
\mathit{H} (\mathit{\hat{Q}} \vert n_B, n_F \rangle)= \mathit{\hat{Q}} (\mathit{H} \vert n_B, n_F \rangle)= (E_B + E_F) (\mathit{\hat{Q}} \vert n_B, n_F \rangle) \:,
\end{equation} 
and this means that the whole energy of the system remains unchanged by the action of the supercharge operators.

\begin{figure} [h!] 
\centering
\begin{tikzpicture} [scale=1.5]
  \draw[gray,very thick] (0,0)--(5,0);
  \draw[gray,very thick,->] (0,0) -- (0,4);

  \draw[ultra thick,blue] (1,0.025)--(2,0.025);
  \draw[ultra thick,blue] (1,1)--(2,1);
  \draw[ultra thick,blue] (1,2)--(2,2);
  \draw[ultra thick,blue] (1,3)--(2,3);

  \draw[black] (1,.22) node[above,left] {${\scriptscriptstyle \mathit{E}_0^{(1)}}$};
  \draw[black] (1,1.1) node[above,left] {${\scriptscriptstyle \mathit{E}_1^{(1)}}$};
  \draw[black] (1,2.1) node[above,left] {${\scriptscriptstyle \mathit{E}_2^{(1)}}$};
  \draw[black] (1,3.1) node[above,left] {${\scriptscriptstyle \mathit{E}_3^{(1)}}$};
  
  \draw[ultra thick,red] (3,1)--(4,1);
  \draw[ultra thick,red] (3,2)--(4,2);
  \draw[ultra thick,red] (3,3)--(4,3);
  
  \draw[black] (4,1.1) node[above,right] {${\scriptscriptstyle \mathit{E}_0^{(2)}}$};
  \draw[black] (4,2.1) node[above,right] {${\scriptscriptstyle \mathit{E}_1^{(2)}}$};
  \draw[black] (4,3.1) node[above,right] {${\scriptscriptstyle \mathit{E}_2^{(2)}}$};
  
  \draw[->] (1.5,3.05) to [bend left] (3.5,3.05);
  \draw[->] (3.5,2.95)  to [bend left] (1.5,2.95);
  \draw[black] (-.1,3.75) node[left] {$\scriptstyle \mathit{E}$};
  \draw[black] (0,0.1) node[above,left] {${\scriptscriptstyle \mathit{E}=0}$};
  \draw[black] (1.5,-.3) node[] {${\scriptstyle \mathit{n}_F = 0 }$};
  \draw[black] (3.5,-.3) node[] {${\scriptstyle \mathit{n}_F = 1 }$};
  \draw[black] (2.5,3.5) node[] {${\scriptscriptstyle \mathit{Q}}$};
  \draw[black] (2.5,2.4) node[] {${\scriptscriptstyle \mathit{Q}^{\dagger}}$};
\end{tikzpicture}
\caption{A schematic view showing the energy levels for a supersymmetric system consisting of only two possible fermionic states $n_F=\{0,1\}$. The supercharge operators $\mathit{Q}$ and $\mathit{Q}^{\dagger}$ exchange bosons and fermions without affecting the energy due to the degeneracy of the energy levels of the two supersymmetric partners except for the ground state of the first partner.}
\label{f2.1} 
\end{figure}
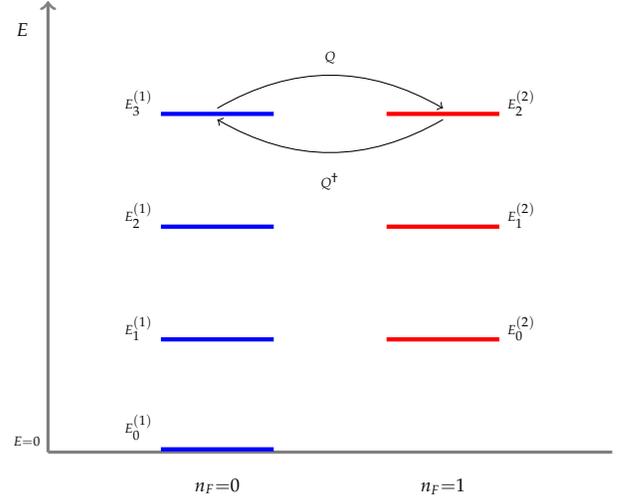

In short, the supercharge operator $\mathit{\hat{Q}}$ acts to change one boson to one fermion leaving the total energy of the system invariant. Conversely, $\mathit{\hat{Q}}^\dagger$ changes a fermion into a boson leaving the energy unchanged. This is illustrated in Figure \ref{f2.1}.
 
 \section{Supersymmetric Quantum Mechanics} \label{Sec2.6}

In this section, we study the general formalism of one-dimensional supersymmetric quantum mechanics. The ideas in this section were discussed in \cite{witten1981dynamical, witten1982constraints, rodrigues2002quantum, ioffe2006susy, gudmundsson2014supersymmetric, berman2002supersymmetric}. We begin by considering a Hamiltonian of the form
\begin{equation} \label{2.26}
\mathit{\hat{H}}= \frac{1}{2} \left( \hat{p}^2 + V(\hat{x}) \right) \mathbbmtt{I}_2 + \frac{1}{2} \hbar \mathit{B}(\hat{x}) \sigma_3 \:,
\end{equation}
where $V(\hat{x})$ is the potential, and $\mathit{B}$ is a magnetic field. If those potential and magnetic fields can be written in terms of some function $\mathit{W}(x)$, as
\begin{equation} \label{2.27}
V(x)= \left( \dfrac{d \mathit{W}(x)}{dx} \right)^2 \equiv {\mathit{W}'}^2 \qquad \qquad \mathit{B}(x)= \left( \dfrac{d^2 \mathit{W}(x)}{dx^2} \right) \equiv \mathit{W}'' \:,
\end{equation}
the Hamiltonian is supersymmetric and we shall refer to the function $\mathit{W}(x)$ as a superpotential. Using Eq. (\ref{2.27}), we can rewrite the Hamiltonian (\ref{2.26}) as
\begin{equation} \label{2.28}
\mathit{\hat{H}}= \frac{1}{2} \left( \hat{p}^2 + {\mathit{W}'}^2 \right) \mathbbmtt{I}_2 + \frac{1}{2} \hbar \sigma_3 \mathit{W}'' \:. 
  \end{equation}
  
At this stage, we define the following two Hermitian supercharge operators:
\begin{align} \label{2.29}
\mathit{\hat{Q}}_1 &= \frac{1}{2} \left( \sigma_1 \hat{p}+\sigma_2 \mathit{W}'(\hat{x}) \right), \nonumber \\
\mathit{\hat{Q}}_2 &= \frac{1}{2} \left( \sigma_2 \hat{p} - \sigma_1 \mathit{W}'(\hat{x}) \right) \:.
\end{align}  
The Hamiltonian (\ref{2.28}) together with the supercharges (\ref{2.29}) constitutes a superalgebra. It is a simple matter of algebra to verify the conditions for the superalgebra given by Eq. (\ref{2.24}). It may be useful before verifying these conditions to check the following commutation relation:
\begin{align}
[ \hat{p}, f(x)] \: g &= [ -i\hbar \frac{\partial }{\partial x}, f(x)] \: g = -i \hbar \left( \frac{\partial }{\partial x} (fg) + f \frac{\partial }{\partial x} (g) \right) \nonumber \\ 
&= -i \hbar \left( f \cdot \frac{\partial g }{\partial x} + \frac{\partial f }{\partial x} \cdot g - f \cdot \frac{\partial g }{\partial x}  \right) = -i\hbar f'(x)\: g \:.
\end{align}
Removing $g$ gives us the commutator of the momentum $\hat{p}$ with an arbitrary function of the position coordinate $x$:
\begin{equation} \label{d2.107}
[ \hat{p}, f(x)] = -i\hbar f'(x).
\end{equation}
Using this relation, we find the following:
\begin{align} \label{d2.108}
[\hat{p}, {\mathit{W}'}^2]&= \hat{p} {\mathit{W}'}^2 - {\mathit{W}'}^2 \hat{p} = - 2i\hbar \mathit{W}' \mathit{W}'', \nonumber \\
[ \mathit{W}', \hat{p}^2 ] &= \mathit{W}' \hat{p}^2- \hat{p}^2 \mathit{W}' = i \hbar \{ \mathit{W}'', \hat{p}\} , \nonumber \\
\{ \hat{p}, \mathit{W}''\} &= \hat{p} \mathit{W}'' + \mathit{W}'' \hat{p} \:.
\end{align}
Now, we can use the outcome of Eq. (\ref{d2.108}) to check the outcome of the commutator of the supercharge $\mathit{\hat{Q}}_1$ with the Hamiltonian $\mathit{\hat{H}}$:
\begin{align}  \label{2.30}
&[\mathit{\hat{Q}}_1, \mathit{\hat{H}}] = \mathit{\hat{Q}}_1 \mathit{\hat{H}} - \mathrm{H} \mathit{\hat{Q}}_1 \nonumber \\ 
& =\resizebox{.28 \textwidth}{!}  {$\frac{1}{4} \left( \sigma_1 \hat{p} + \sigma_2 \mathit{W}' \right) \cdot \left( \left( \hat{p}^2 + {\mathit{W}'}^2\right) \mathbbmtt{I}_2 + \hbar \mathit{W}'' \sigma_3 \right)$} \nonumber \\  
& \quad \resizebox{.28 \textwidth}{!}  {$- \frac{1}{4} \left( \left(\hat{p}^2 + {\mathit{W}'}^2\right) \mathbbmtt{I}_2 + \hbar \mathit{W}'' \sigma_3 \right) \cdot \left(\sigma_1 \hat{p} + \sigma_2 \mathit{W}'\right)$} \nonumber \\
& = \resizebox{.44 \textwidth}{!} {$\frac{1}{4}  \left( \sigma_1 \hat{p}^3 + \sigma_1 \hat{p} {\mathit{W}'}^2 + \sigma_1 \sigma_3 \hbar \hat{p} \mathit{W}''  + \sigma_2 \mathit{W}' \hat{p}^2 + \sigma_2 \mathit{W}' {\mathit{W}'}^2 +\sigma_2 \sigma_3 \hbar \mathit{W}'  \mathit{W}''\right) $}\nonumber \\
 & \quad - \resizebox{.44 \textwidth}{!} {$ \frac{1}{4} \left( \sigma_1 \hat{p}^3 +\sigma_2 \hat{p}^2 \mathit{W}' + \sigma_1 {\mathit{W}'}^2 \hat{p} +\sigma_2  {\mathit{W}'}^2 \mathit{W}'+ \sigma_3 \sigma_1 \hbar \mathit{W}'' \hat{p} +\sigma_3 \sigma_2 \hbar \mathit{W}'' \mathit{W}'\right) $} \nonumber \\ 
& = \resizebox{.4 \textwidth}{!} {$ \frac{1}{4} \left( \sigma_1 [\hat{p}, {\mathit{W}'}^2] - \sigma_2 i \hbar \{\hat{p}, \mathit{W}'' \} + \sigma_2 [\mathit{W}', \hat{p}] + 2 \sigma_1 i \hbar \mathit{W}' \mathit{W}''  \right)$} \nonumber \\
& = 0 \:.
    \end{align}
Similarly, one can confirm that
\begin{equation} \label{2.31}
[\mathit{\hat{Q}}_2, \mathit{\hat{H}}] = \mathit{\hat{Q}}_2 \mathit{\hat{H}} - \mathit{\hat{H}} \mathit{\hat{Q}}_2 = 0. 
\end{equation}
Furthermore, we can find that
\begin{align} \label{2.32}
\{ \hat{Q}_1, \hat{Q}_1 \} & = \hat{Q}_1 \hat{Q}_1 + \hat{Q}_1 \hat{Q}_1 = 2 \hat{Q}_1^2 \nonumber \\
& = \frac{1}{2} \left( \sigma_1 \hat{p} + \sigma_2 \mathit{W}' \right)^2 \nonumber \\ 
& =  \frac{1}{2} \left(\sigma_1^2 \hat{p}^2 + \sigma_2^2 {\mathit{W}'}^2 
+\sigma_1  \sigma_2 \hat{p} \mathit{W}' + \sigma_2 \sigma_1 \mathit{W}' \hat{p} \right) \nonumber \\
&= \frac{1}{2} \left( ( \hat{p}^2 +  {\mathit{W}'}^2 ) \mathbbmtt{I}_2 + i \sigma_3  ( \hat{p} \mathit{W}' - \mathit{W}' \hat{p}) \right) \nonumber \\ 
&= \frac{1}{2} \left( ( \hat{p}^2 + {\mathit{W}'}^2 )\mathbbmtt{I}_2 + \hbar \sigma_3 \mathit{W}'' \right)  \equiv \mathrm{H} \:,
\end{align}
and  similarly, we have
\begin{equation} \label{2.34}
\{ \mathit{\hat{Q}}_2^{\dagger}, \mathit{\hat{Q}}_2\} =\hat{\mathrm{H}} \:. 
\end{equation}
Note that, to do the previous calculations, we used the Pauli matrices properties: $\sigma_1^2=\sigma_2^2=\sigma_3^2= \mathbbmtt{I}_2$ and $\sigma_1 \sigma_2=-\sigma_2 \sigma_1= i \sigma_3$.
It is clear from Eqs. (\ref{2.30},\ref{2.31}, \ref{2.32}, and \ref{2.34}) that the Hamiltonian (\ref{2.28}) together with the supercharges (\ref{2.29}) satisfies the conditions for the superalgebra given by Eq. (\ref{2.24}).

Indeed, if we define the quantities $\mathit{Q}$ and $\mathit{Q}^{\dagger}$   as
\begingroup \allowdisplaybreaks
\begin{align}
\mathit{\hat{Q}} &= \mathit{\hat{Q}}_1 -i \mathit{\hat{Q}}_2 \nonumber \\
&= \frac{1}{2} \left( \sigma_1 \hat{p} + \sigma_2 \mathit{W}' \right) - \frac{1}{2} i \left( \sigma_2 \hat{p} - \sigma_1 \mathit{W}' \right) \nonumber \\
&= \frac{1}{2} \left( (\sigma_1 - i \sigma_2) \hat{p} + (\sigma_2+i \sigma_1) \mathit{W}'\right)  \nonumber \\
&= \frac{1}{2} (\sigma_1 - i \sigma_2) \left( \hat{p} + i \mathit{W}' \right) \nonumber \\
&= \sigma_{-} \left( \hat{p} +i \mathit{W}' \right) \:, 
\end{align} \endgroup
and
\begin{equation}
\hat{Q}^\dagger = \hat{Q}_1 +i \hat{Q}_2 = \sigma_{+} \left( \hat{p} -i \mathit{W}' \right) \:, 
\end{equation}
where $\sigma_{-}$ and $\sigma_{+}$, respectively, are defined as
\begin{align}
\sigma_{-} &= \frac{1}{2} (\sigma_1 - i \sigma_2) = \left( \begin{matrix}0&0\\1&0 \end{matrix} \right), \nonumber \\
\sigma_{+} &= \frac{1}{2} (\sigma_1 + i \sigma_2) = \left( \begin{matrix}0&1\\0&0 \end{matrix} \right) \:,
\end{align}
it is straightforward to check that the Hamiltonian $\mathit{\hat{H}}$ can be expressed as
\begin{equation} \label{f2.116}
\mathit{\hat{H}} = \{ \mathit{\hat{Q}}, \mathit{\hat{Q}}^{\dagger} \} \:,
\end{equation}  
and it commutes with both the operators $\mathit{\hat{Q}}$ and $\mathit{\hat{Q}}^{\dagger}$:
\begin{equation} \label{2.39}
[\mathit{\hat{Q}}, \mathit{\hat{H}}]=0 \quad \& \quad [\mathit{\hat{Q}}^{\dagger}, \mathit{\hat{H}}]=0 \:.
\end{equation}
Furthermore, one also finds
\begin{equation}
\{\mathit{\hat{Q}}, \mathit{\hat{Q}}\}=0 \quad \& \quad \{\mathit{\hat{Q}}^{\dagger}, \mathit{\hat{Q}}^{\dagger}\}=0 \:.
\end{equation}

As we have seen in the case of the supersymmetric Harmonic oscillator, a supersymmetric system with two independent supercharges, with the possible exception of the energy of the ground eigenstate, all the energy levels are split into two eigenstates with either $n_F=0$ or $n_F=1$. For a spinor quantum mechanical system, this implies that the excited energy eigenstates come in degenerate spin-up/spin-down pairs $\vert E_n, \uparrow \rangle$/$\vert E_n, \downarrow \rangle$. These degenerate spin-up/spin-down pairs are related to the acts of the supercharge operators $\mathit{\hat{Q}}$ and $\mathit{\hat{Q}}^{\dagger}$, where the supercharge operators $\mathit{\hat{Q}}$ convert the degenerate spin-up state to the degenerate spin-down state without making any change in the energy eigenvalue of the states:
\begin{equation}
\hat{Q} \vert E_n , \uparrow \rangle \sim \left( \begin{matrix} 0 &0 \\ 1&0 \end{matrix} \right) \left( \begin{matrix} \uparrow \\ 0  \end{matrix} \right) = \left( \begin{matrix} 0 \\ \downarrow  \end{matrix} \right) \:,
\end{equation}  
while the supercharge operators $\mathit{\hat{Q}}^{\dagger}$ convert the degenerate spin-down state to the degenerate spin-up state without making any change in the energy eigenvalue of the states:
\begin{equation}
\hat{Q}^\dagger \vert E_n , \downarrow \rangle \sim \left( \begin{matrix} 0 &1 \\ 0&0 \end{matrix} \right) \left( \begin{matrix} 0 \\ \downarrow  \end{matrix} \right) = \left( \begin{matrix} \uparrow \\ 0  \end{matrix} \right) \:.
\end{equation}

To sum up, the supersymmetric transformations occur due to the supercharge operators $\mathit{\hat{Q}}$ and $\mathit{\hat{Q}}^\dagger$. In this case, it causes transforms between the energy eigenstates spin-up/spin-down which have the same energy eigenvalues.

\section{Supersymmetric Ground State} \label{Sec2.7}

So far, we have investigated all the required information to describe a supersymmetric quantum mechanical system. In this section, we study the supersymmetric quantum mechanical ground state. The argument in this section follows from \cite{argyres1996introduction, argyres2001introduction, guilarte2006n, junker2012supersymmetric, salomonson1982fermionic}. Let us now write the expression of the supersymmetric Hamiltonian (\ref{2.28}) as the summation of two separate terms: a Hamiltonian $\mathit{\hat{H}}_{+}$ and a Hamiltonian $\mathit{\hat{H}}_{-}$, then, we have
\begin{equation}
\mathit{\hat{H}}_{\pm} = \frac{1}{2} \left( \hat{p}^2 + {\mathit{W}'}^2 \right) \pm \frac{1}{2} \hbar \mathit{W}'' \:.
\end{equation}     
In the eigenbasis of $\sigma_3$, the supersymmetric Hamiltonian is diagonal:
\begin{equation} \label{f2.122}
\mathit{\hat{H}} \equiv \left( \begin{matrix}  \mathit{\hat{H}}_{+} & 0 \\ 0 & \mathit{\hat{H}}_{-}  \end{matrix} \right) = \left( \begin{matrix}  \mathit{A}^{\dagger} \mathit{A} & 0 \\ 0 & \mathit{A} \mathit{A}^{\dagger}  \end{matrix} \right) \:, 
\end{equation} 
and the supercharge operators $\mathit{\hat{Q}}$ and $\mathit{\hat{Q}}^{\dagger}$ can be written as
\begin{equation} \label{f2.123}
\mathit{\hat{Q}} = \left( \begin{matrix}  0 & 0 \\ \mathit{A} & 0 \end{matrix} \right), \qquad 
\mathit{\hat{Q}}^{\dagger} = \left( \begin{matrix}  0 & \mathit{A}^{\dagger} \\ 0 & 0  \end{matrix} \right) \:,
\end{equation}
where
\begin{align} \label{f2.124}
\mathit{A} &= \hat{p} + i \mathit{W}', \qquad
\mathit{A}^{\dagger} = \hat{p} - i \mathit{W}' \:.
\end{align}

According to Eq. (\ref{f2.116}), we can write the supersymmetric Hamiltonian $\mathit{\hat{H}}$ in terms of the supercharges operators $\mathit{\hat{Q}}$ and $\mathit{\hat{Q}}^{\dagger}$ as follows:
\begin{align} \label{f2.125}
\mathit{\hat{H}} &= \mathit{\hat{Q}} \mathit{\hat{Q}}^{\dagger} + \mathit{\hat{Q}}^{\dagger} \mathit{\hat{Q}}  
= 2 {\mathit{\hat{Q}}}^2 = 2 \mathit{\hat{Q}}^{\dagger^2} \:.
\end{align}
In the last step, we used the fact that the supercharges are Hermitian operators. We see from Eq. (\ref{f2.125}) that the Hamiltonian $\mathit{\hat{H}}$ can be written in terms of the squares of the supercharge operators. For this reason, the energy of any eigenstate of this Hamiltonian must be positive or zero. Let us now consider that $\vert \Psi_0 \rangle$ is the ground state of the supersymmetric Hamiltonian $\mathit{\hat{H}}$. Based on Eq. (\ref{f2.125}), it can be seen that the ground state can have a zero energy only if it satisfies the following two conditions:
\begin{align} \label{f2.126}
& \resizebox{.39 \textwidth}{!} {$ \mathit{E}_0 = \langle \Psi_0 \vert \mathit{\hat{H}} \vert \Psi_0 \rangle = \langle \Psi_0 \vert {\mathit{\hat{Q}}}^2 \vert \Psi_0 \rangle = 0 \quad  \Longrightarrow \quad \mathit{\hat{Q}} \vert \Psi_0 \rangle =0, $} \nonumber \\
& \resizebox{.39 \textwidth}{!}  {$\mathit{E}_0 = \langle \Psi_0 \vert \mathit{\hat{H}} \vert \Psi_0 \rangle = \langle \Psi_0 \vert \mathit{\hat{Q}}^{\dagger^2} \vert \Psi_0 \rangle = 0 \qquad  \Longrightarrow \quad \mathit{\hat{Q}}^{\dagger} \vert \Psi_0 \rangle =0. $} 
\end{align}
Therefore, if there exists a state which is annihilated by each of the supercharge operators $\mathit{\hat{Q}}$ and $\mathit{\hat{Q}}^{\dagger}$ which means that it is invariant under the supersymmetry transformations, such a state is automatically the zero-energy ground state. However, on the other hand, any state that is not invariant under the supersymmetry transformations has a positive energy. Thus, if there is a supersymmetric state, it is the zero-energy ground state and it is said that the supersymmetry is unbroken. 

Moreover, since the supersymmetry algebra (\ref{2.39}) implies that the supercharge operators $\mathit{\hat{Q}}$ and $\mathit{\hat{Q}}^{\dagger}$ commute with the supersymmetric  Hamiltonian $\mathit{\hat{H}}$, so all the eigenstates of $\mathit{\hat{H}}$ are doubly degenerate. For that reason, it will be convenient to write the supersymmetric ground state of the system in terms of two components, $\psi_{0}^{\pm}$, as follows:
\begin{equation} \label{f2.127}
\Psi_{0}= \left( \begin{matrix}  \psi_{0}^{+} \\  \psi_{0}^{-} \end{matrix} \right) \:. 
\end{equation} 
One can now solve eigenvalue problem $\mathit{\hat{Q}} \vert \psi_0 \rangle =0$ to find the zero-energy ground-state wave function. If we make substitution using Eqs. (\ref{f2.123},  \ref{f2.127}), we get
\begin{align}
\mathit{\hat{Q}} \vert \psi_{0} \rangle =
\left( \begin{matrix}  0 & 0 \\ \mathit{A} & 0 \end{matrix} \right) \left( \begin{matrix}  \psi_{0}^{+}  \\ \psi_{0}^{-} \end{matrix} \right) =0 \:,
\end{align} 
and then, if we make substitution using Eq. (\ref{f2.124}), our problem reduces to solve the first-order differential equation:
\begin{equation}
\mathit{A} \vert \psi_{0}^{+} \rangle  =\left( \hat{p} + i \mathit{W}' \right) \vert \psi_{0}^{+} \rangle = 0 \quad \Longrightarrow \quad
- i\hbar \dfrac{\partial \psi_{0}^{+}}{\partial x} + i \mathit{W}' \psi_{0}^{+} = 0 \:.
\end{equation}
It is simple and straightforward to solve the previous differential equation as follows"
\begin{equation}
\dfrac{d \psi_{0}^{+}}{\psi_{0}^{+}} = \dfrac{\mathit{W}'}{\hbar} dx \quad \Longrightarrow \quad
\ln \psi_{0}^{+} = \frac{\mathit{W}'}{\hbar} + \ln A \:.
\end{equation}
Thus,
\begin{equation}
\psi_{0}^{+} = A e^{\frac{\mathit{W}}{\hbar}} \:,
\end{equation}
and similarly,
\begin{equation}
\psi_{0}^{+} = B e^{- \frac{\mathit{W}}{\hbar}} \:.
\end{equation}
Now, the general form of the ground-state wave function may be expressed as
\begin{equation} \label{f2.133}
\Psi_0 = \left( \begin{matrix} A e^{\frac{\mathit{W}}{\hbar}} \\ B e^{- \frac{\mathit{W}}{\hbar}}. \end{matrix} \right)\:.
\end{equation}
 
In fact, it has no physical meaning to find the ground-state wave function $\psi_0$ if there are no normalizable solutions of such form. So we need now to normalize the ground-state wave function (\ref{f2.133}) by determining the values of the two constants $A$ and $B$. The ground-state wave function $\psi_0(x)$ to be normalizable must vanish at the positive and the negative infinite $x$-values. In order to satisfy this condition, we have to set $\vert \mathit{W}(x) \vert \rightarrow \infty$ as $\vert x \vert \rightarrow \infty$. Then, we have the following three possible cases for the supersymmetric ground-state wave function:
 
\begin{itemize}
\item[1.] 
The first case is when $\mathit{W}(x) \rightarrow + \infty$ as $x\rightarrow \pm \infty$. In this case, the superpotential $\mathit{W}(x)$ is an even function and it is positive at the boundaries. Figure \ref{F2.1} describes the behavior of the potential $\mathit{V}(x)$ in this case. Since $\mathit{W}(x)$ is positive, we cannot normalize the wave function $\psi_{0}^{+} = \mathit{A} e^{\frac{\mathit{W}(x)}{\hbar}}$, but we can choose $\mathit{A}=0$. However, we can normalize the wave function $\varphi_{0}^{-} = B e^{-\frac{\mathit{W}(x)}{\hbar}}$ to find the value of the constant $B$. The complete ground-state wave function, in this case, may be expressed in the general form
\begin{equation}
\Psi_0 = \left( \begin{matrix} 0 \\ B e^{-\frac{\mathit{W}(x)}{\hbar}} \end{matrix} \right) \:.
\end{equation} 
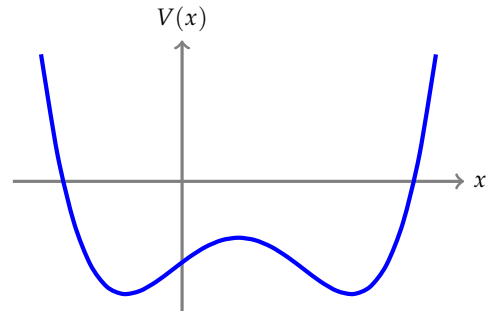
\begin{figure} [h!]
\centering
\begin{tikzpicture} [scale=1.5]
      \draw[very thick, gray,->] (-2,0) -- (2,0) node[right,black] {$x$};
      \draw[very thick, gray,->] (-.5,-1.15) -- (-.5,1.25) node[above,black] {$V(x)$};
      \draw[domain=-1.75:1.75,smooth,variable=\x,blue,ultra thick] plot ({\x},{.5*\x*\x*\x*\x-\x*\x-.5});
    \end{tikzpicture} 
\caption{The even superpotential $\mathit{W}(x)$ is an even function, where $\mathit{W}(x) \rightarrow + \infty$ as $x \rightarrow \pm \infty$.}
\label{F2.1} 
\end{figure}

\item[2.] 
The second case is when $\mathit{W}(x) \rightarrow - \infty$ as $x\rightarrow \pm \infty$. In this case, the superpotential $\mathit{W}(x)$ is an even function and it is negative at the boundaries. The behavior of the potential $\mathit{V}(x)$ in this case is described in Figure \ref{F2.2}. Since $\mathit{W}(x)$ is negative, we can normalize the wave function $\Psi_0^a = A e^{\frac{\mathit{W}(x)}{\hbar}} \nonumber $ and find the value of the constant $A$; however we cannot normalize the wave function $\Psi_0^b = B e^{-\frac{\mathit{W}(x)}{\hbar}}$, but we can choose $B=0$.
The complete ground-state wave function, in this case, shall be expressed in the general form
\begin{equation}
\Psi_0 = \left( \begin{matrix} A e^{\frac{\mathit{W}(x)}{\hbar}} \\ 0 \end{matrix} \right) \:.
\end{equation}
\begin{figure} [h!]
\centering
\begin{tikzpicture} [scale=1.5]
      \draw[very thick, gray,->] (-2,0) -- (2,0) node[right,thick,black] {$x$};
      \draw[very thick, gray,->] (.5,-1.15) -- (.5,1.25) node[above,black, thick] {$V(x)$};
      \draw[domain=-1.75:1.75,smooth,variable=\x,blue,ultra thick] plot ({\x},{-.5*\x*\x*\x*\x+\x*\x+.5});
    \end{tikzpicture}    
\caption{The superpotential$\mathit{W}(x)$ is an even function, where $\mathit{W}(x) \rightarrow - \infty$ as $x \rightarrow \pm \infty$.} 
\label{F2.2} 
\end{figure}
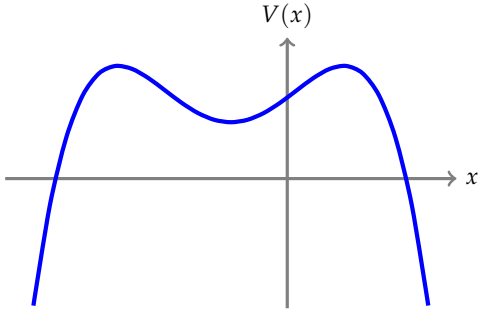

\item[3.] 
The other two cases correspond to $\mathit{W}(x) \rightarrow \{+ \infty, - \infty \}$ as $x\rightarrow \{+ \infty, - \infty\}$, and  $\mathit{W}(x) \rightarrow \{- \infty, + \infty \}$ as $x\rightarrow \{+ \infty, - \infty\}$. In those two cases, the superpotential $\mathit{W}(x)$ is an odd function and it is positive at one of the boundaries and negative at the other boundary. The behavior of the potential $\mathit{V}(x)$ in this case is described in Figure \ref{F2.3}. In those two cases, both the two constants $A$ and $B$ vanish and we cannot normalize the wave function. Therefore. we have
\begin{equation}
\Psi_0 = 0.
\end{equation}
Thus, since those two forms of the zero-energy ground-state wave function cannot be normalized, this means that they do not exist.
\begin{figure} [h!]
\centering
\begin{subfigure}[b]{.4\textwidth}
\centering
\begin{tikzpicture} [scale=.9]
\draw[very thick, gray,->] (-2.8,.75) -- (2.8,.75) node[right,black] {$x$};
\draw[very thick, gray,->] (-.25,-2) -- (-.25,2.25) node[above,black] {$V(x)$};
\draw[domain=-2.595:2.595,smooth,variable=\x,blue,ultra thick] plot ({\x},{.1*\x*\x*\x*\x*\x-.708*\x*\x*\x+\x});\end{tikzpicture} 
\caption{$\mathit{W}(x) \rightarrow - \infty $ as $x \rightarrow  - \infty$, and $\mathit{W}(x) \rightarrow  \infty$ as $x \rightarrow   \infty$.}
\end{subfigure} 
\space
\begin{subfigure}[b]{.4\textwidth}
\centering
\begin{tikzpicture} [scale=.9]
\draw[very thick, gray,->] (-2.8,-.75) -- (2.8,-.75) node[right,black] {$x$};
\draw[very thick, gray,->] (-.25,-2) -- (-.25,2.25) node[above,black] {$V(x)$};
\draw[domain=-2.595:2.595,smooth,variable=\x,blue,ultra thick] plot ({\x},{-.1*\x*\x*\x*\x*\x+.708*\x*\x*\x-\x});
\end{tikzpicture} 
\caption{$\mathit{W}(x) \rightarrow  \infty$ as $x \rightarrow  - \infty$, and $\mathit{W}(x) \rightarrow  - \infty$ as $x \rightarrow   \infty$.}
\end{subfigure}
\caption{The superpotential$\mathit{W}(x)$ is an odd function.}
\label{F2.3}   
\end{figure}
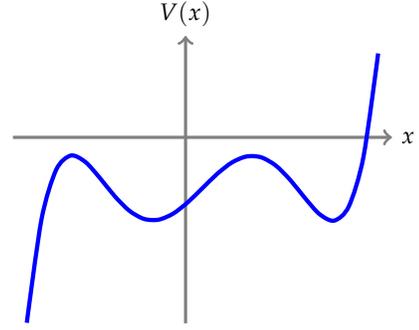
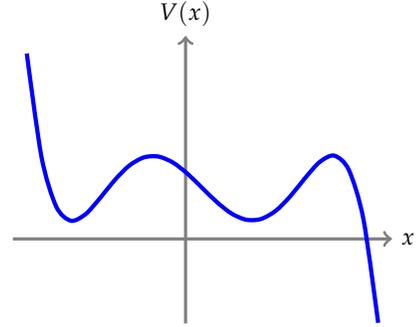 
\end{itemize}

In short, the zero-energy ground-state wave function can be normalized when the superpotential $\mathit{W}(x)$ has an even number of zeros. In this case, there exists a zero-energy ground state which is the true vacuum state and the supersymmetry is unbroken.  However on the other hand, if the superpotential $\mathit{W}(x)$ has an odd number of zeros, the zero-energy ground-state wave function cannot be normalized. Then, we immediately realize that there is no zero-energy ground state in such case and the supersymmetry is spontaneously broken.

\section{Corrections to the Ground-State Energy} \label{Sec.2.8}

In this section, we consider the example of the supersymmetry harmonic oscillator which has a unique zero-energy ground state. Then, we use the conventional perturbation theory to examine whether it has nonvanishing corrections to the energy of the ground state, and accordingly, it can be responsible for the spontaneous breaking of the supersymmetry. To this end, in the next two subsections, based on \cite{goldstein1965classical,
griffiths2005introduction, book:17492, sakurai2011modern,
shankar2012principles},  we are going to compute both the first and the second corrections to the ground-state energy of the supersymmetric harmonic oscillator. For instance, let us recall the general form of the supersymmetric quantum mechanics Hamiltonian, which is given by Eq. (\ref{2.28}): 
  \begin{equation}\label{2.62}
\mathit{\hat{H}} = \frac{1}{2} (\hat{p}^2 + \mathit{W}'^2) \mathbbmtt{I}_2 + \frac{1}{2} \hbar \mathit{W}'' \sigma_3.
  \end{equation}  
Consider now that the superpotential $\mathit{W}(x)$ is defined as
\begin{equation} \label{q2.138}
\mathit{W}(q)= \frac{1}{2} \omega \hat{q}^2 + \mathit{g} \hat{q}^3 \quad \Rightarrow \quad \begin{matrix}  \;\;\; \mathit{W}'  =  \omega \hat{q}+ 3 \mathit{g} \hat{q}^2 \\ \mathit{W}''  = \omega + 6 \mathit{g} \hat{q} \:, \end{matrix} 
\end{equation}
where $\mathit{g}$ is a perturbation. If $\mathit{g}=0$, the previous Hamiltonian reduces to the supersymmetric harmonic oscillator Hamiltonian:
\begin{equation}   
\mathit{\hat{H}}^{0} = \frac{1}{2} (\hat{p}^2 + \omega^2 q^2) \mathbbmtt{I}_2 + \frac{1}{2} \hbar \omega \sigma_3 \:.
\end{equation} 
We have already solved this Hamiltonian and found that it has a unique zero-energy ground state,
\begin{equation}
\mathit{E}_0^{(0)}=0 \:,
\end{equation}
and we verify that this state is invariant under supersymmetry. Based on our argument in the previous section and after normalization, the unbroken supersymmetric ground-state wave function can be written as follows:
\begin{equation} \label{f2.142}
\Psi_0^{(0)}= \left( \begin{matrix} 0  \\ (\frac{\omega}{\pi \hbar})^{\frac{1}{4}} e^{- \frac{\omega q^2}{2 \hbar}} \end{matrix} \right) \:.
\end{equation}

Now, when $\mathit{g}$ is small, the Hamiltonian (\ref{2.62}) can be written as
\begin{equation}
\mathit{\hat{H}}= \mathit{\hat{H}}^{0} + \mathit{\hat{H}}' \:,
\end{equation}
where $\mathit{\hat{H}}^{0}$ is the Hamiltonian of the unperturbed system and $\mathit{\hat{H}}'$ a perturbation:
\begin{equation}
\mathit{\hat{H}}'= (\frac{9}{2} \mathit{g}^2 \hat{q}^4 + 3\omega \mathit{g} \hat{q}^3) \mathbbmtt{I}_2 + 3 \hbar \mathit{g} \hat{q} \sigma_3 \:.
\end{equation}
Moreover, if we consider the matrix representation, the perturbative Hamiltonian $\mathit{\hat{H}}$ acting on the ground-state wave function $\Psi_0^{(0)}$ yields
\begin{equation}
\resizebox{.48 \textwidth}{!}  {$\mathit{\hat{H}}' \Psi_0^{(0)} = \left( \begin{matrix} \frac{9}{2} \mathit{g}^2 \hat{q}^4 + 3\omega \mathit{g} \hat{q}^3 + 3 \hbar \mathit{g} \hat{q} & 0 \\ 0 & \frac{9}{2} \mathit{g}^2 \hat{q}^4 + 3\omega \mathit{g} \hat{q}^3 - 3 \hbar \mathit{g} \hat{q} \end{matrix} \right) \left( \begin{matrix} 0  \\ (\frac{\omega}{\pi \hbar})^{\frac{1}{4}} e^{- \frac{\omega q^2}{2 \hbar}} \end{matrix} \right) =0.$}
\end{equation}
In view of this last equation, it is clear that for this combination the impact of the Hamiltonian $\mathit{\hat{H}}'$ on the wave function $\Psi_0^{(0)}$ is only due to the component
\begin{equation}
\mathcal{\hat{H}}'= \frac{9}{2} \mathit{g}^2 \hat{q}^4 + 3\omega \mathit{g} \hat{q}^3 - 3 \hbar \mathit{g} \hat{q} \:.
\end{equation}
As a result, we see that it is enough to consider the Hamiltonian $\mathcal{\hat{H}}'$ to compute the first and second corrections to the ground-state energy (see \ref{AppendixB.1}).

Furthermore, we have to mention here that, for the supersymmetric harmonic oscillator, except for the ground state, all the energy levels are degenerate to two energy states. However, because we are just interested in computing the correction to the ground-state energy, which is not degenerate, we can use the nondegenerate perturbation theory. For more explanation about this point see \ref{AppendixB.1}.

In addition, it is useful to remember from quantum mechanics that the wave function of degree $n$ can be obtained by the following recursion formula:
  \begin{align} \label{2.60}
\hat{q} \psi_n &= \sqrt{\frac{(n+1)\hbar}{2 \omega}} \psi_{n+1} + \sqrt{\frac{n \hbar}{2 \omega}} \psi_{n-1} \:.
  \end{align}     
As well, it is important to recall the following relation:
\begin{equation}
\langle \psi_n \vert \psi_m \rangle = \delta_{nm} \:.
\end{equation}
We are ready now to move forward to the next two subsections and compute the first and the second corrections to the energy of the ground state. 

\subsection{The First-Order Correction}

We know from the previous argument that the ground state of the supersymmetric Hamiltonian is nondegenerate. Therefore, we can use the time-independent nondegenerate perturbation theory to compute the first-order correction to the ground-state energy of the supersymmetric harmonic oscillator as follows:
\begin{align} \label{2.34}
\mathit{E}_0^{(1)} & = \langle \psi_0^{(0)} \vert \mathcal{\hat{H}}' \vert \psi_0^{(0)} \rangle \nonumber \\
& = \langle \psi_0^{(0)} \vert \frac{9}{2} \mathit{g}^2 \hat{q}^4 + 3\omega \mathit{g} \hat{q}^3 - 3 \hbar \mathit{g} \hat{q} \vert \psi_0^{(0)} \rangle  \nonumber  \\ 
& = \resizebox{.4 \textwidth}{!}  {$\frac{9}{2} \mathit{g}^2 \langle \psi_0^{(0)} \vert \hat{q}^4 \vert \psi_0^{(0)} \rangle  + 3\omega \mathit{g} \langle \psi_0^{(0)} \vert \hat{q}^3\vert \psi_0^{(0)} \rangle  - 3 \hbar \mathit{g} \langle \psi_0^{(0)} \vert \hat{q} \vert \psi_0^{(0)} \rangle \:. $}
\end{align}
The terms $\hat{q}^3 \psi_0$ and $\hat{q} \psi_0$ are linearly independent of $\psi_0$ so that both the second and the third terms of Eq. (\ref{2.34}) are zero. However,
\begin{equation}
\hat{q}^4 \psi_0 = \frac{3\hbar^4}{2\omega^4}\psi_4 + \frac{7\hbar^4}{4\omega^4}\psi_2+\frac{9\hbar^4}{16\omega^4}\psi_0 \:,
\end{equation}    
then, substituting it into Eq. (\ref{2.34}) gives us,
\begin{align}\label{2.71}
\mathit{E}_0^{(1)} & = \frac{9}{2} \mathit{g}^2 \langle \psi_0^{(0)} \vert \hat{q}^4 \vert \psi_0^{(0)} \rangle = \frac{9}{2} g^2 \times \frac{3}{4} \frac{\hbar^2}{\omega^2} \langle \psi_0^{(0)} \vert \psi_0^{(0)} \rangle  \nonumber \\
&= \frac{27}{8} \frac{\hbar^2}{\omega^2} \mathit{g}^2 \simeq  \mathcal{O}(g^2) \:.  
\end{align}
In the view of the last equation, the first-order correction to the ground-state energy reduces to zero. Another method with more details to calculations of the first-order correction of the ground-state energy is given in \ref{AppendixB.1}.

\subsection{The Second-Order Correction}

From the nondegenerate time-independent perturbation theory, the second-order correction to the energy is given by
\begin{equation}\label{2.72}
\mathit{E}_2^{(0)} = \sum_{m \neq n} \frac{\vert \langle \psi_n^{(0)} \vert \mathcal{\hat{H}}' \vert \psi_m^{(0)} \rangle \vert^2}{\mathit{E}_m^{(0)} - \mathit{E}_n^{(0)}} \:.
\end{equation}
Since the perturbation contains only terms of $q$, $q^3$, and $q^4$, the numerator of Eq. (\ref{2.72}) is zero for all $m$ values except $m=1,3,4$. For more explanation about this point, see \ref{AppendixB.2}. Therefore, the second-order correction to ground-state energy of the supersymmetric harmonic oscillator is computed as follows:
\begin{align}\label{2.73}
\mathit{E}_2^{(0)} & = \frac{\vert \langle \psi_0^{(0)} \vert \frac{9}{2} \mathit{g}^2 \hat{q}^4 + 3 \omega \mathit{g} \hat{q}^3 - 3 \hbar \mathit{g} \hat{q} \vert \psi_1^{(0)} \rangle \vert^2}{E_0^{(0)} - \mathit{E}_1^{(0)}}  \nonumber \\
& {\quad} +  \frac{\vert \langle \psi_0^{(0)} \vert \frac{9}{2} \mathit{g}^2 \hat{q}^4 + 3 \omega \mathit{g} \hat{q}^3 - 3 \hbar \mathit{g} \hat{q}  \vert \psi_2^{(0)} \rangle \vert^2}{\mathit{E}_0^{(0)} - \mathbb{E}_2^{(0)}}
\nonumber \\
&{\quad} + \frac{\vert \langle \psi_0^{(0)} \vert \frac{9}{2} \mathit{g}^2 \hat{q}^4 + 3 \omega \mathit{g} \hat{q}^3 - 3 \hbar \mathit{g} \hat{q}  \vert \psi_3^{(0)} \rangle \vert^2}{\mathit{E}_0^{(0)} - \mathit{E}_3^{(0)}} \nonumber \\
&{\quad}+  \frac{\vert \langle \psi_0^{(0)} \vert \frac{9}{2} \mathit{g}^2 \hat{q}^4 + 3 \omega \mathit{g} \hat{q}^3 - 3 \hbar \mathit{g} \hat{q}  \vert \psi_4^0 \rangle \vert^2}{\mathit{E}_0^{(0)} - \mathit{E}_4^{(0)}} \:.
\end{align}
We can simplify the previous equation by calculating the numerator of each term using the recurrence relation for the harmonic oscillator energy eigenfunction given by Eq. (\ref{2.60}). We get
\begin{align} \label{2.74}
\hat{q} \psi_1 &= \sqrt{\frac{\hbar}{\omega}} \psi_2 + \sqrt{\frac{\hbar}{2 \omega}} \psi_0, \nonumber \\
\hat{q}^3 \psi_1 &= \sqrt{\frac{3 \hbar^3}{\omega^3}} \psi_4+ 3 \sqrt{\frac{\hbar^3}{\omega^3}} \psi_2+ \sqrt{\frac{9 \hbar^3}{8 \omega^3}} \psi_0, \nonumber \\
\hat{q}^3 \psi_3 &= \resizebox{.35 \textwidth}{!}  {$\sqrt{\frac{15 \hbar^3}{\omega^3}} \psi_6 + \sqrt{\frac{72 \hbar^3}{\omega^3}} \psi_4+ \sqrt{\frac{243 \hbar^3}{8 \omega^3}} \psi_2 + \sqrt{\frac{3 \hbar^3}{4 \omega^3}} \psi_0,$} \nonumber \\
\hat{q}^4 \psi_2 &= \resizebox{.38 \textwidth}{!}  {$\sqrt{\frac{45\hbar^4}{2\omega^4}} \psi_6 + \sqrt{\frac{147\hbar^4}{\omega^4}} \psi_4 + 12.5 \sqrt{\frac{\hbar^4}{\omega^4}} \psi_2 + \sqrt{\frac{9\hbar^4}{2\omega^4}} \psi_0,$}  \nonumber \\
\hat{q}^4 \psi_4 &= \resizebox{.42 \textwidth}{!}  {$\sqrt{\frac{105 \hbar^4}{\omega^4}} \psi_8 + \sqrt{\frac{945\hbar^4}{4\omega^4}} \psi_6 + \sqrt{\frac{2725\hbar^4}{16\omega^4}} \psi_4 + \sqrt{\frac{81\hbar^4}{2\omega^4}} \psi_2 + \sqrt{\frac{3\hbar^4}{2\omega^4}} \psi_0 \:.$}
\end{align} 
  
In Eq. (\ref{2.74}), we only list the terms which contain $\psi_0$. The other terms do not affect our calculations, since all of them give us zero (see \ref{AppendixB.1}). Using Eq. (\ref{2.74}) to calculate Eq. (\ref{2.73}), we get
\begin{align} \label{2.75}
\mathit{E}_2^{(0)} & = - \frac{\mathit{g}^2}{\hbar \omega} \left( 3\omega \times \frac{9 \hbar^3}{\omega^3} - 3 \hbar \times \sqrt{\frac{\hbar}{2\omega}} \right)^2
- \frac{81 \mathit{g}^4}{8\hbar \omega} \left( \sqrt{\frac{9\hbar^4}{2 \omega^4}} \right)^2 \nonumber \\
&{\quad}-\frac{g^2}{3 \hbar \omega} \left( 3\omega \times \sqrt{\frac{3\hbar^3}{4\omega^3}} \right)^2 - \frac{\mathit{g}^4}{4\hbar \omega} \left( \frac{9}{2} \times \sqrt{\frac{3 \hbar^4}{2 \omega^4}} \right)^2 \nonumber \\
& = - \frac{9}{8} \frac{\hbar^2}{\omega^2} \mathit{g}^2 - \frac{729}{16} \frac{\hbar^3}{\omega^5} \mathit{g}^4 - \frac{9}{4} \frac{\hbar^2}{\omega^2} \mathit{g}^2 - \frac{243}{32} \frac{\hbar^3}{\omega^5} \mathit{g}^4 \nonumber \\
& = - \frac{27}{8} \frac{\hbar^2}{\omega^2} \mathit{g}^2  - \frac{1701}{32} \frac{\hbar^3}{\omega^5} \mathit{g}^4 \:.
\end{align}   
For more details, another method to compute the second-order correction of the ground-state energy is given in \ref{AppendixB.2}.

Finally, with regard to Eqs. (\ref{2.71} and \ref{2.75}), we realize that up to the second order the energy corrections to the ground-state energy vanish. This is concluded as
\begin{align}
\mathcal{O}(g)&=0, \nonumber \\
\mathcal{O}(g^2)& = \frac{27}{8}\frac{\hbar^2}{\omega^2} - \frac{27}{8}\frac{\hbar^2}{\omega^2}=0 \:.
\end{align}  
It is worth noting that the second term in Eq. (\ref{2.75}) should be canceled when we extend the calculation to the fourth-order perturbation corrections. Showing this would make the calculations here more complicated and confusing. However,  the result can be expanded to any finite order in perturbation theory. Hence, there are no corrections to the energy of the ground state and the supersymmetry remains unbroken at any finite order of perturbation theory.

\section{Properties of SUSY Quantum Mechanics} \label{Sec.2.9}

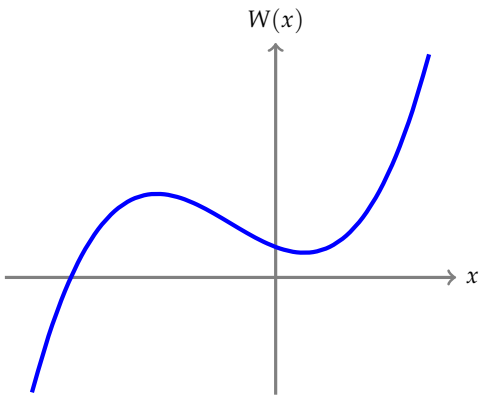
\begin{figure} [h!]
\centering
\begin{tikzpicture} [scale=1.2]
      \draw[very thick, gray,->] (-2.5,-0.6) -- (2.5,-0.6) node[right,black] {$x$};
      \draw[very thick, gray,->] (.5,-1.9) -- (.5,2) node[above,black] {$\mathit{W}(x)$};
      \draw[domain=-2.2:2.2,smooth,variable=\x,blue,ultra thick] plot ({\x},{.3*\x*\x*\x-.6*\x});
    \end{tikzpicture} 
\caption{The superpotential of the harmonic oscillator ground state.}
\label{F2.4} 
\end{figure}
A supersymmetric quantum mechanics system is said to have unbroken supersymmetry if it has a zero-energy ground state, which is $\mathit{E}_0=0$,  while if the system has a positive ground-state energy, $\mathit{E}_0>0$, it is said to have a broken supersymmetry \cite{engbrant2012supersymmetric}. For example, in Section \ref{Sec2.5}, we have studied the Hamiltonian of the supersymmetric harmonic oscillator and showed that it obeys the superalgebra. Then, we calculated the ground-state energy for that system and found evidence that it vanishes, $E_0=0$, and consequently, this supersymmetric harmonic oscillator has an unbroken supersymmetry.

In the previous section, we applied a small perturbation $\mathit{g}$ to the supersymmetric harmonic oscillator and calculated the effect on the ground-state energy. We found out that the perturbation does not affect the energy of the ground state at the second order in perturbation theory, but this result can be expanded to any finite order. This means that the supersymmetry breaking does not occur because of perturbation, and it must be due to the nonperturbative effects.

Again, let us consider the same potential Eq. (\ref{q2.138}), which we have used before in Section \ref{Sec.2.8}:
\begin{equation}
W(q)= \frac{1}{2} \omega \hat{q}^2 + g \hat{q}^3 \:.
\end{equation}  
As we discussed in Section \ref{Sec2.7}, the wave function of the ground state should have three zeros, $\Psi_0 \rightarrow +\infty$ as $x \rightarrow +\infty$, and $\Psi_0 \rightarrow -\infty$ as $x \rightarrow -\infty$. Figure \ref{F2.4}, shows an arbitrary diagram presenting how the wave function of the ground state should look like, and Eq. (\ref{2.78}) gives us the nonnormalized form of the ground-state wave function:
\begin{equation}\label{2.78}
\Psi_0 = \left( \begin{matrix} A e^{-\frac{W}{\hbar}} \\ B e^{-\frac{W}{\hbar}} \end{matrix} \right) \:.
\end{equation}
Acutely, Eq. (\ref{2.78}) could not be normalized, and the only way to solve the Schr\"{o}dinger equation for the ground state is taking $\Psi_0=0$. This means that the wave function of the ground state is not excited, even if the perturbation theory told us that it is excited and has no energy correction.

Thus, the perturbation technique gives us incorrect results for both the wave function and the energy spectrum and fails to give an explanation to the supersymmetry breaking.

\section{Conclusions}

In this study, we studied the basic aspects of supersymmetric quantum mechanics. We started with introducing the algebra of Grassmann variables and then looked into quantum mechanics of the supersymmetric harmonic oscillator, which includes fermionic as well as bosonic fields. Afterward, we investigated the algebraic structure of supersymmetric quantum mechanics. We started by investigating the superalgebra using Dirac brackets. Then, we introduced the concept of the supercharge operators $\hat{\mathit{Q}}$ and $\hat{\mathit{Q}}^\dagger$. In general, the supersymmetry is constructed by introducing supersymmetric transformations which are generated by the supercharge operators, where the role of the supercharges is to change the bosonic state into the fermionic state and vice versa, while the Lagrangian remains invariant. Moreover, we have presented the basic properties of supersymmetric quantum mechanics.
 
Furthermore, we illustrated, for a supersymmetric quantum mechanical system, that the energy spectrum is degenerate except for the ground state, which must have a zero eigenvalue in order for the system to have an unbroken supersymmetry. Also, we have explained that if there is a supersymmetric state, it is the zero-energy ground state. If such a zero-energy ground state exists, it is said that the supersymmetry is unbroken. So far, there has been no unbroken supersymmetry observed in nature, and if nature is described by supersymmetry, of course, it must be broken.

In fact, supersymmetry may be broken spontaneously at any order of perturbation theory or dynamically due to nonperturbative effects. To examine this statement, we studied the normalization of the ground state of the supersymmetric harmonic oscillator. Then, we used perturbation theory to calculate the corrections to the ground-state energy. We found out that the perturbation does not affect the energy of the ground state at second order in perturbation theory, but this result can be expanded to any finite order. This means that the supersymmetry breaking is not seen in perturbation theory, and it must be due to the nonperturbative effects.
 
\section*{Acknowledgements}

The author acknowledges the research office of the University of the Witwatersrand and the African Institute for Mathematical Sciences (Ghana) for financial support. 

\appendix

\section{The Generalized Hamilton's Equations of Motion} \label{AppendixA.1}

Recall the Hamiltonian (\ref{a2.31}):
\begin{equation} \label{A.1}
\mathit{H} = \frac{1}{2} \left( p^2 + \omega^2 q^2 \right) - i \omega \psi_1 \psi_2 \:.
\end{equation}
We showed in Subsection \ref{SubSec.2.3.3} that the previous Hamiltonian could be written in the form (\ref{a2.36}):
\begin{equation} \label{A.2}
\mathit{H} = \frac{1}{2} \left( p^2 + \omega^2 q^2 \right) +4 i \omega \pi^1 \pi^2 \:.
\end{equation}
Also consider the momenta $\pi^{\alpha}$ and the primary constraint $\phi^{\alpha}$,  respectively, define as
\begin{align} \label{A.3}
\pi^{\alpha} &= \dfrac{\partial \mathit{L}}{\partial \dot{\psi}_{\alpha}} = - \frac{i}{2} \delta^{\alpha \beta} \psi_{\beta}, \nonumber \\
\phi^{\alpha} &= \pi^{\alpha} + \frac{i}{2} \delta^{\alpha \beta} \psi_{\beta} \:.
\end{align}
Downward, we explain how to use the Poisson bracket to get the generalized Hamilton's equations of motion, which we have only write down in Eq. (\ref{b2.44}).
\begin{itemize}
\item[(i)]
The extended Hamiltonian (\ref{a2.41}) is written as
\begin{equation}
\mathit{H} = \dot{q}_i p^i + \dot{\psi}_{\alpha} \pi^{\alpha} - \phi^A \lambda_A - \mathit{L} \:.
\end{equation}
Therefore,
\begin{equation}
\dfrac{\partial \mathrm{H}}{\partial p^i} = \dot{q}_i - \dfrac{\partial \phi^A}{\partial p^i} \lambda_A \:.
\end{equation}
Thus,
\begin{equation} \label{A.6}
\dot{q}_i = \dfrac{\partial \mathrm{H}}{\partial p}  + \dfrac{\partial \phi^A}{\partial p} \lambda_A \:.
\end{equation}
Using Poisson brackets, we find that
\begin{align} \label{A.7}
\resizebox{.11 \textwidth}{!}  {$\{ q_i, \mathit{H}+ \phi^A \lambda_A \}_P$} & = \resizebox{.30 \textwidth}{!}  {$\left( \dfrac{\partial q_i}{ \partial q_i} \dfrac{ \partial (\mathit{H} + \phi^A \lambda_A)}{ \partial p^i} - \dfrac{ \partial (\mathit{H} + \phi^A \lambda_A)}{ \partial q_i} \dfrac{\partial q_i}{\partial p^i} \right)$} \nonumber \\ 
& \quad \resizebox{.30 \textwidth}{!}  {$- \left( \dfrac{\partial q_i}{\partial \psi_\alpha} \dfrac{\partial (\mathit{H} + \phi^A \lambda_A)}{\partial \pi^\alpha} - \dfrac{\partial (\mathit{H} + \phi^A \lambda_A)}{\partial \psi_\alpha} \dfrac{\partial q_i}{\partial \pi^\alpha} \right)$} \nonumber \\ 
& = \resizebox{.08 \textwidth}{!}  {$\dfrac{\partial \mathit{H}}{\partial p^i} + \dfrac{\partial \pi^A}{\partial p^i} \lambda_A. $}
\end{align} 
From Eqs. (\ref{A.6}, \ref{A.7}), we obtain
\begin{equation}
\dot{q}_i = \dfrac{\partial \mathrm{H}}{\partial p}  + \dfrac{\partial \phi^A}{\partial p} \lambda_A = \{ q_i, \mathrm{H}+ \phi^A \lambda_A \}_P \:.
\end{equation}
\item[(ii)]
\begin{align}
\dfrac{\partial \mathit{H}}{\partial q_i} & = - \dfrac{\partial \mathit{L}}{\partial q_i}- \dfrac{ \partial \phi^A}{\partial q_i} \lambda_A  
= - \dfrac{d}{dt} \dfrac{\partial \mathit{L}}{\partial \dot{q_i}}- \dfrac{ \partial \phi^A}{\partial q_i} \lambda_A \nonumber \\
& = - \dot{p}^i- \dfrac{ \partial \phi^A}{\partial q_i} \lambda_A \:. 
\end{align}
Thus,
\begin{equation}\label{A.10}
\dot{p}^i = - \dfrac{\partial \mathit{H}}{\partial q_i} - \dfrac{ \partial \phi^A}{\partial q_i} \lambda_A \:.
\end{equation}
Using the Poisson brackets, we find that
\begin{align} \label{A.11}
\resizebox{.11 \textwidth}{!}  {$\{ p^i, \mathit{H}+ \phi^A \lambda_A \}_P$} &= \resizebox{.30 \textwidth}{!}  {$\left( \dfrac{\partial p^i}{ \partial q_i} \dfrac{ \partial (\mathit{H} + \phi^A \lambda_A)}{ \partial p^i} - \dfrac{ \partial (\mathit{H} + \phi^A \lambda_A)}{ \partial q_i} \dfrac{\partial p^i}{\partial p^i} \right)$} \nonumber \\ 
& \quad \resizebox{.30 \textwidth}{!}  {$- \left( \dfrac{\partial p^i}{\partial \psi_\alpha} \dfrac{\partial (\mathrm{H} + \phi^A \lambda_A)}{\partial \pi^\alpha} - \dfrac{\partial (\mathit{H} + \phi^A \lambda_A)}{\partial \psi_\alpha} \dfrac{\partial p^i}{\partial \pi^\alpha} \right)$} \nonumber \\
& = \resizebox{.08 \textwidth}{!}  {$- \dfrac{\partial \mathrm{H}}{\partial q_i} - \dfrac{\partial \phi^A}{\partial q_i}.$}
\end{align}
From Eqs. (\ref{A.10}, \ref{A.11}), we obtain
\begin{equation}
\dot{p}^i = - \dfrac{\partial \mathit{H}}{\partial q_i} - \dfrac{ \partial \phi^A}{\partial q_i} \lambda_A = \{ p^i, \mathit{H}+ \phi^A \lambda_A \}_P \:.
\end{equation}
\item[(iii)]
\begin{align}
\dfrac{\partial \mathit{H}}{\partial \pi^\alpha} & = - \dot{\psi_\alpha}- \dfrac{ \partial \phi^A}{\partial q\pi^\alpha} \lambda_A \:. 
\end{align}
Thus,
\begin{equation} \label{A.14}
\dot{\psi_\alpha} = - \dfrac{\partial \mathit{H}}{\partial \pi^\alpha} - \dfrac{ \partial \phi^A}{\partial q\pi^\alpha} \lambda_A \:.
\end{equation}
Using the Poisson brackets, we find that
\begin{align} \label{A.15}
\resizebox{.11 \textwidth}{!}  {$ \{ \psi_\alpha, \mathit{H}+ \phi^A \lambda_A \}_P$} & = \resizebox{.30 \textwidth}{!}  {$\left( \dfrac{\partial \psi_\alpha}{ \partial q_i} \dfrac{ \partial (\mathit{H} + \phi^A \lambda_A)}{ \partial p^i} - \dfrac{ \partial (\mathit{H} + \phi^A \lambda_A)}{ \partial q_i} \dfrac{\partial p^i}{\partial \psi_\alpha} \right)$}  \nonumber \\ 
& \quad \resizebox{.30 \textwidth}{!}  {$- \left( \dfrac{\partial \psi_\alpha}{\partial \psi_\alpha} \dfrac{\partial (\mathit{H} + \phi^A \lambda_A)}{\partial \pi^\alpha} - \dfrac{\partial (\mathit{H} + \phi^A \lambda_A)}{\partial \psi_\alpha} \dfrac{\partial \psi_\alpha }{\partial \pi^\alpha} \right)$}  \nonumber \\
& = \resizebox{.08 \textwidth}{!}  {$- \dfrac{\partial \mathit{H}}{\partial \pi^\alpha} - \dfrac{\partial \phi^A}{\partial \pi^\alpha} \:.$}
\end{align}
From Eqs. (\ref{A.14}, \ref{A.15}), we obtain
\begin{equation}
\dot{\psi_\alpha} = - \dfrac{\partial \mathit{H}}{\partial \pi^\alpha} - \dfrac{ \partial \phi^A}{\partial q\pi^\alpha} \lambda_A = \{ \psi_\alpha, \mathit{H}+ \phi^A \lambda_A \}_P \:.
\end{equation}
\item[(iv)]
\begin{align}
\dfrac{\partial \mathit{H}}{\partial \psi_\alpha} & = - \dfrac{\partial \mathit{L}}{\partial \psi_\alpha}- \dfrac{ \partial \phi^A}{\partial q \psi_\alpha} \lambda_A \nonumber \\
&= - \dfrac{d}{dt} \left(\dfrac{\partial \mathit{L}}{\partial \dot{\psi}_\alpha}\right) - \dfrac{ \partial \phi^A}{\partial \psi_\alpha} \lambda_A \nonumber \\
& = - \dot{\pi}^\alpha - \dfrac{ \partial \phi^A}{\partial \psi_\alpha} \lambda_A \:.
\end{align}
Thus,
\begin{equation} \label{A.18}
\dot{\pi}^{\alpha} = - \dfrac{\partial \mathit{H}}{\partial \psi_{\alpha}} - \dfrac{ \partial \phi^A}{\partial \psi_\alpha} \lambda_A \:.
\end{equation}
Using the Poisson brackets, we obtain
\begin{align} \label{A.19}
\resizebox{.11 \textwidth}{!}  {$ \{ \pi_\alpha, \mathit{H}+ \phi^A \lambda_A \}_P$} &= \resizebox{.30 \textwidth}{!}  {$ \left( \dfrac{\partial \pi_\alpha}{ \partial q_i} \dfrac{ \partial (\mathit{H} + \phi^A \lambda_A)}{ \partial p^i} - \dfrac{ \partial (\mathit{H} + \phi^A \lambda_A)}{ \partial q_i} \dfrac{\partial p^i}{\partial \pi_\alpha} \right)$} \nonumber \\ 
& \quad \resizebox{.30 \textwidth}{!}  {$- \left( \dfrac{\partial \pi_\alpha}{\partial \psi_\alpha} \dfrac{\partial (\mathit{H} + \phi^A \lambda_A)}{\partial \pi^\alpha} - \dfrac{\partial (\mathit{H} + \phi^A \lambda_A)}{\partial \psi_\alpha} \dfrac{\partial \pi_\alpha }{\partial \pi^\alpha} \right)$} \nonumber \\
&=  \resizebox{.08 \textwidth}{!}  {$- \dot{\pi}^\alpha - \dfrac{ \partial \phi^A}{\partial \psi_\alpha} \lambda_A \:.$}
\end{align}
From Eqs. (\ref{A.18}, \ref{A.19}), we obtain
\begin{equation}
\dot{\pi}^{\alpha} = - \dfrac{\partial \mathit{H}}{\partial \psi_{\alpha}} - \dfrac{ \partial \phi^A}{\partial \psi_\alpha} \lambda_A = \{ \pi_\alpha, \mathit{H}+ \phi^A \lambda_A \}_P \:. 
\end{equation}
\item[(v)]
By definition from Eq. (\ref{A.3}),
\begin{align}
\phi^A &= \pi^A + \frac{i}{2} \psi_A \nonumber \\
& = - \frac{i}{2} \psi_A + \frac{i}{2} \psi_A \nonumber \\
&= 0 \:.
\end{align}
\end{itemize}

\section{Dirac Bracket and the Superalgebra} \label{AppendixA.2}

From (\ref{AppendixA.1}), recall the expressions of the Hamiltonian (\ref{A.1}) and the extended Hamiltonian (\ref{A.2}):
\begin{equation} \label{A.22}
\mathit{H} = \frac{1}{2} \left( p^2 + \omega^2 q^2 \right) - i \omega \psi_1 \psi_2.
\end{equation}
As well, take into account  the  for the constraint (\ref{A.3}) and suppose that $\{ \alpha ,\beta\}$ can only take the values $\{1,2\}$; then, we have
\begin{align} \label{A.23}
\phi^1 &= \pi^1 + \frac{i}{2} \psi_1, \nonumber \\
\phi^2 &= \pi^2 + \frac{i}{2} \psi_2 \:. 
\end{align}  
Furthermore, consider the supercharge formula, and once again suppose that $\{ \alpha ,\beta\}$ can only take the values $\{1,2\}$, then we have 
\begin{align}\label{A.24}
\mathit{Q}_\alpha  = p \psi_\alpha + \omega q \epsilon_{\alpha \beta} \delta^{\beta \gamma} \psi_\gamma \quad \Rightarrow \quad 
\begin{split} 
\mathit{Q}_1  &= p \psi_1 + \omega q \psi_2  \\
\mathit{Q}_2  &= p \psi_2 - \omega q \psi_1 \:,
\end{split}
\end{align}
where $\epsilon_{\alpha \beta}$ and $\delta^{\alpha \beta}$ are the Levi-Civita symbol and Kronecker delta function, respectively.

To verify that the supercharges (\ref{A.24}) together with the Hamiltonian (\ref{A.22}) and the constraints (\ref{A.23}) satisfy the Dirac bracket superalgebra (\ref{aa2.73}), we need as a first step to calculate the Dirac brackets of the supercharges:
\begin{align} \label{A.25}
& \resizebox{.42 \textwidth}{!}  {$ \{Q_\alpha, Q_\beta \}_D = \{Q_1, Q_1\}_D + \{Q_1, Q_2 \}_D + \{Q_2, Q_1\}_D + \{Q_2, Q_2 \}_D \:.$}
\end{align}
To simplify the calculation of Eq. (\ref{A.25}), let us start with calculating the Poisson bracket of the supercharges and the constraints.
\begin{align} \label{A.26}
\resizebox{.07 \textwidth}{!}  {$ \{Q_1,Q_1\}_P $} &= \resizebox{.16 \textwidth}{!}  {$ \left( \frac{\partial Q_1 }{\partial q_i} \frac{\partial Q_1 }{\partial p^i} - \frac{\partial Q_1 }{\partial p^i} \frac{\partial Q_1}{\partial q_i} \right)$} \nonumber \\
& \quad \resizebox{.35 \textwidth}{!}  {$- \left( \frac{\partial Q_1}{\partial \psi_1} \frac{\partial Q_1}{\partial \pi^1} + \frac{\partial Q_1}{\partial \pi^1} \frac{\partial Q_1}{\partial \psi_1} \right)
- \left( \frac{\partial Q_1}{\partial \psi_2} \frac{\partial Q_1}{\partial \pi^2} + \frac{\partial Q_1}{\partial \pi^2} \frac{\partial Q_1}{\partial \psi_2} \right)$} \nonumber \\
& = 0 \:,
\end{align}
\begin{align}
\resizebox{.07 \textwidth}{!}  {$\{Q_1,Q_2\}_P$} & =\resizebox{.16 \textwidth}{!}  {$ \left( \frac{\partial Q_1 }{\partial q_i} \frac{\partial Q_2 }{\partial p^i} - \frac{\partial Q_1 }{\partial p^i} \frac{\partial Q_2}{\partial q_i} \right)$} \nonumber \\
& \quad \resizebox{.35 \textwidth}{!}  {$- \left( \frac{\partial Q_1}{\partial \psi_1} \frac{\partial Q_2}{\partial \pi^1} + \frac{\partial Q_1}{\partial \pi^1} \frac{\partial Q_2}{\partial \psi_1} \right)
- \left( \frac{\partial Q_1}{\partial \psi_2} \frac{\partial Q_2}{\partial \pi^2} + \frac{\partial Q_1}{\partial \pi^2} \frac{\partial Q_2}{\partial \psi_2} \right)$} \nonumber \\
& =  \omega \left( \psi_1^2 + \psi_2^2 \right) \:,
\end{align}
\begin{align}
\resizebox{.07 \textwidth}{!}  {$ \{Q_2,Q_1\}_P$} &= \resizebox{.16 \textwidth}{!}  {$ \left( \frac{\partial Q_2 }{\partial q_i} \frac{\partial Q_1 }{\partial p^i} - \frac{\partial Q_2 }{\partial p^i} \frac{\partial Q_1}{\partial q_i} \right)$} \nonumber \\
& \quad \resizebox{.35 \textwidth}{!}  {$- \left( \frac{\partial Q_2}{\partial \psi_1} \frac{\partial Q_1}{\partial \pi^1} + \frac{\partial Q_2}{\partial \pi^1} \frac{\partial Q_1}{\partial \psi_1} \right)
- \left( \frac{\partial Q_2}{\partial \psi_2} \frac{\partial Q_1}{\partial \pi^2} + \frac{\partial Q_2}{\partial \pi^2} \frac{\partial Q_1}{\partial \psi_2} \right)$} \nonumber \\
& = - \omega \left(\psi_1^2 + \psi_2^2 \right) \:,
\end{align}
\begin{align}
\resizebox{.07 \textwidth}{!}  {$\{Q_2,Q_2\}_P$} &= \resizebox{.16 \textwidth}{!}  {$\left( \frac{\partial Q_2 }{\partial q_i} \frac{\partial Q_2 }{\partial p^i} - \frac{\partial Q_2 }{\partial p^i} \frac{\partial Q_2}{\partial q_i} \right)$} \nonumber \\
& \quad \resizebox{.35 \textwidth}{!}  {$- \left( \frac{\partial Q_2}{\partial \psi_1} \frac{\partial Q_2}{\partial \pi^1} + \frac{\partial Q_2}{\partial \pi^1} \frac{\partial Q_2}{\partial \psi_1} \right)
- \left( \frac{\partial Q_2}{\partial \psi_2} \frac{\partial Q_2}{\partial \pi^2} + \frac{\partial Q_2}{\partial \pi^2} \frac{\partial Q_2}{\partial \psi_2} \right)$} \nonumber \\
&= 0 \:,
\end{align}
\begin{align}
\resizebox{.07 \textwidth}{!}  {$\{Q_1, \phi^1\}_P$} &= \resizebox{.16 \textwidth}{!}  {$ \left( \frac{\partial Q_1 }{\partial q_i} \frac{\partial \phi^1 }{\partial p^i} - \frac{\partial Q_1 }{\partial p^i} \frac{\partial \phi^1}{\partial q_i} \right)$} \nonumber \\
& \quad \resizebox{.35 \textwidth}{!}  {$- \left( \frac{\partial Q_1}{\partial \psi_1} \frac{\partial \phi^1}{\partial \pi^1} + \frac{\partial Q_1}{\partial \pi^1} \frac{\partial \phi^1}{\partial \psi_1} \right)
- \left( \frac{\partial Q_1}{\partial \psi_2} \frac{\partial \phi^1}{\partial \pi^2} + \frac{\partial Q_1}{\partial \pi^2} \frac{\partial \phi^1}{\partial \psi_2} \right)$} \nonumber \\
&= - p \:,
\end{align}
\begin{align}
\resizebox{.07 \textwidth}{!}  {$\{\phi^1, Q_1\}_P$} &= \resizebox{.16 \textwidth}{!}  {$\left( \frac{\partial \phi^1 }{\partial q_i} \frac{\partial Q_1 }{\partial p^i} - \frac{\partial \phi^1 }{\partial p^i} \frac{\partial Q_1}{\partial q_i} \right)$} \nonumber \\
& \quad \resizebox{.35 \textwidth}{!}  {$- \left( \frac{\partial \phi^1}{\partial \psi_1} \frac{\partial Q_1}{\partial \pi^1} + \frac{\partial \phi^1}{\partial \pi^1} \frac{\partial Q_1}{\partial \psi_1} \right)
- \left( \frac{\partial \phi^1}{\partial \psi_2} \frac{\partial Q_1}{\partial \pi^2} + \frac{\partial \phi^1}{\partial \pi^2} \frac{\partial Q_1}{\partial \psi_2} \right)$} \nonumber \\
&= -p \:,
\end{align}
\begin{align}
\resizebox{.07 \textwidth}{!}  {$\{Q_1, \phi^2\}_P$} & = \resizebox{.16 \textwidth}{!}  {$\left( \frac{\partial Q_1 }{\partial q_i} \frac{\partial \phi^2 }{\partial p^i} - \frac{\partial Q_1 }{\partial p^i} \frac{\partial \phi^2}{\partial q_i} \right)$} \nonumber \\
& \quad \resizebox{.35 \textwidth}{!}  {$- \left( \frac{\partial Q_1}{\partial \psi_1} \frac{\partial \phi^2}{\partial \pi^1} + \frac{\partial Q_1}{\partial \pi^1} \frac{\partial \phi^2}{\partial \psi_1} \right)
- \left( \frac{\partial Q_1}{\partial \psi_2} \frac{\partial \phi^2}{\partial \pi^2} + \frac{\partial Q_1}{\partial \pi^2} \frac{\partial \phi^2}{\partial \psi_2} \right)$} \nonumber \\
&= - \omega q \:,
\end{align}
\begin{align}
\resizebox{.07 \textwidth}{!}  {$\{\phi^2 ,Q_1 \}_P$} &= \resizebox{.16 \textwidth}{!}  {$\left( \frac{\partial \phi^2 }{\partial q_i} \frac{\partial Q_1 }{\partial p^i} - \frac{\partial \phi^2 }{\partial p^i} \frac{\partial Q_1}{\partial q_i} \right)$} \nonumber \\
& \quad \resizebox{.35 \textwidth}{!}  {$- \left( \frac{\partial \phi^1}{\partial \psi_2} \frac{\partial Q_1}{\partial \pi^1} + \frac{\partial \phi^2}{\partial \pi^1} \frac{\partial Q_1}{\partial \psi_1} \right)
- \left( \frac{\partial \phi^2}{\partial \psi_2} \frac{\partial Q_1}{\partial \pi^2} + \frac{\partial \phi^2}{\partial \pi^2} \frac{\partial Q_1}{\partial \psi_2} \right)$} \nonumber \\
&= - \omega q \:,
\end{align}
\begin{align}
\resizebox{.07 \textwidth}{!}  {$\{Q_2, \phi^1 \}_P$} &= \resizebox{.16 \textwidth}{!}  {$\left( \frac{\partial Q_2 }{\partial q_i} \frac{\partial \phi^1 }{\partial p^i} - \frac{\partial Q_2 }{\partial p^i} \frac{\partial \phi^1}{\partial q_i} \right)$} \nonumber \\
& \quad \resizebox{.35 \textwidth}{!}  {$- \left( \frac{\partial Q_2}{\partial \psi_1} \frac{\partial \phi^1}{\partial \pi^1} + \frac{\partial Q_2}{\partial \pi^1} \frac{\partial \phi^1}{\partial \psi_1} \right)
- \left( \frac{\partial Q_2}{\partial \psi_2} \frac{\partial \phi^1}{\partial \pi^2} + \frac{\partial Q_2}{\partial \pi^2} \frac{\partial \phi^1}{\partial \psi_2} \right)$} \nonumber \\
&=  \omega q \:,
\end{align}
\begin{align}
\resizebox{.07 \textwidth}{!}  {$ \{\phi^1,Q_2 \}_P$} &= \resizebox{.16 \textwidth}{!}  {$\left( \frac{\partial \phi^1 }{\partial q_i} \frac{\partial Q_2 }{\partial p^i} - \frac{\partial \phi^1 }{\partial p^i} \frac{\partial Q_2}{\partial q_i} \right)$} \nonumber \\
& \quad \resizebox{.35 \textwidth}{!}  {$- \left( \frac{\partial \phi^1}{\partial \psi_1} \frac{\partial Q_2}{\partial \pi^1} + \frac{\partial \phi^1}{\partial \pi^1} \frac{\partial Q_2}{\partial \psi_1} \right)
- \left( \frac{\partial \phi^1}{\partial \psi_2} \frac{\partial Q_2}{\partial \pi^2} + \frac{\partial \phi^1}{\partial \pi^2} \frac{\partial Q_2}{\partial \psi_2} \right)$} \nonumber \\
&=  \omega q \:,
\end{align}
\begin{align}
\resizebox{.07 \textwidth}{!}  {$\{Q_2, \phi^2 \}_P$} &= \resizebox{.16 \textwidth}{!}  {$\left( \frac{\partial Q_2 }{\partial q_i} \frac{\partial \phi^2 }{\partial p^i} - \frac{\partial Q_2 }{\partial p^i} \frac{\partial \phi^2}{\partial q_i} \right)$} \nonumber \\
& \quad \resizebox{.35 \textwidth}{!}  {$- \left( \frac{\partial Q_2}{\partial \psi_1} \frac{\partial \phi^2}{\partial \pi^1} + \frac{\partial Q_2}{\partial \pi^1} \frac{\partial \phi^2}{\partial \psi_1} \right)
- \left( \frac{\partial Q_2}{\partial \psi_2} \frac{\partial \phi^2}{\partial \pi^2} + \frac{\partial Q_2}{\partial \pi^2} \frac{\partial \phi^2}{\partial \psi_2} \right)$} \nonumber \\
&= - p \:,
\end{align}
\begin{align} \label{A.37}
\resizebox{.07 \textwidth}{!}  {$\{\phi^2,Q_2 \}_P$} &= \resizebox{.16 \textwidth}{!}  {$\left( \frac{\partial \phi^2 }{\partial q_i} \frac{\partial Q_2 }{\partial p^i} - \frac{\partial \phi^2 }{\partial p^i} \frac{\partial Q_2}{\partial q_i} \right)$} \nonumber \\
& \quad \resizebox{.35 \textwidth}{!}  {$- \left( \frac{\partial \phi^2}{\partial \psi_1} \frac{\partial Q_2}{\partial \pi^1} + \frac{\partial \phi^2}{\partial \pi^1} \frac{\partial Q_2}{\partial \psi_1} \right)
- \left( \frac{\partial \phi^2}{\partial \psi_2} \frac{\partial Q_2}{\partial \pi^2} + \frac{\partial \phi^2}{\partial \pi^2} \frac{\partial Q_2}{\partial \psi_2} \right)$} \nonumber \\
&= - p \:.
\end{align}

Using Dirac bracket definition and Eq.s ( \ref{A.26}-\ref{A.37}), then, we get
\begin{align} \label{A.38}
& \resizebox{.08 \textwidth}{!}  {$\{Q_1, Q_1 \}_D$} =\resizebox{.28 \textwidth}{!}  {$ \{Q_1,Q_1\}_P - \{Q_1,\phi^A \}_P C_{AB} \{ \phi^B, Q_1\}_P$} \nonumber \\
&= \resizebox{.45 \textwidth}{!}  {$\{Q_1,Q_1\}_P - \{Q_1,\phi^1 \}_P C_{11} \{ \phi^1, Q_1\}_P - \{Q_1,\phi^2 \}_P C_{22} \{ \phi^2, Q_1\}_P$} \nonumber \\
&= - i ( p^2 + \omega^2 q^2) \:,
\end{align}
\begin{align} \label{A.39}
& \resizebox{.08 \textwidth}{!}  {$\{Q_1, Q_2 \}_D$} = \resizebox{.28 \textwidth}{!}  {$\{Q_1,Q_2 \}_P - \{Q_1,\phi^A \}_P C_{AB} \{ \phi^B, Q_2 \}_P$} \nonumber \\
&=  \resizebox{.45\textwidth}{!}  {$\{Q_1,Q_2\}_P - \{Q_1,\phi^1 \}_P C_{11} \{ \phi^1, Q_2\}_P - \{Q_1,\phi^2 \}_P C_{22} \{ \phi^2, Q_2\}_P$} \nonumber \\
&= \omega (\psi_1^2 + \psi_2^2) \:,
\end{align}
\begin{align} \label{A.40}
& \resizebox{.08 \textwidth}{!}  {$\{Q_2, Q_1 \}_D $}= \resizebox{.28 \textwidth}{!}  {$\{Q_2,Q_1\}_P - \{Q_2,\phi^A \}_P C_{AB} \{ \phi^B, Q_1\}_P$} \nonumber \\
&=  \resizebox{.45 \textwidth}{!}  {$\{Q_2,Q_1\}_P - \{Q_2,\phi^1 \}_P C_{11} \{ \phi^1, Q_1\}_P - \{Q_2,\phi^2 \}_P C_{22} \{ \phi^2, Q_1\}_P$}  \nonumber \\
&= - \omega (\psi_1^2 + \psi_2^2) \:,
\end{align}
\begin{align} \label{A.41}
& \resizebox{.08 \textwidth}{!}  {$\{Q_2, Q_2 \}_D$} = \resizebox{.28 \textwidth}{!}  {$\{Q_2,Q_2 \}_P - \{Q_2,\phi^A \}_P C_{AB} \{ \phi^B, Q_2 \}_P$} \nonumber \\
&= \resizebox{.45 \textwidth}{!}  {$\{Q_2,Q_2\}_P - \{Q_2,\phi^1 \}_P C_{11} \{ \phi^1, Q_2\}_P - \{Q_2,\phi^2 \}_P C_{22} \{ \phi^2, Q_2\}_P$} \nonumber \\
&= - i ( p^2 + \omega^2 q^2) \:.
\end{align}
Substituting Eqs. (\ref{A.38}, \ref{A.39}, \ref{A.40}, and \ref{A.41})  into Eq. (\ref{A.25}), thus, we find
\begin{align} \label{A.42}
\resizebox{.08 \textwidth}{!}  {$\{Q_\alpha, Q_\beta \}_D$} &= \resizebox{.38 \textwidth}{!}  {$\{Q_1, Q_1 \}_D + \{Q_1, Q_2 \}_D + \{Q_2, Q_1 \}_D + \{Q_2, Q_2 \}_D$} \nonumber \\
& = \resizebox{.38 \textwidth}{!}  {$- i ( p^2 + \omega^2 q^2) + \omega (\psi_1^2 + \psi_2^2) - \omega (\psi_1^2 + \psi_2^2) - i ( P^2 + \omega^2 q^2)$} \nonumber \\
&= - 2i ( P^2 + \omega^2 q^2) \:.
\end{align}
On the other hand,
\begin{align} \label{A.43}
-2 i \delta_{\alpha \beta} \mathit{H} &=  -2i \delta_{11} \mathrm{H} - 2i \delta_{22} \mathit{H} \nonumber \\ 
&= - 2i ( p^2 + \omega^2 q^2) - 2i ( P^2 + \omega^2 q^2) \nonumber \\
&= - 2i ( p^2 + \omega^2 q^2) \:. 
\end{align}
Eqs. (\ref{A.42}) and (\ref{A.43}) imply that
\begin{equation} \label{A.44}
\{Q_\alpha, Q_\beta \}_D = -2i \delta_{\alpha \beta} \mathit{H} \:.
\end{equation}

So far,  as a second step to verify that the supercharges (\ref{A.24}) together with the Hamiltonian (\ref{A.22}) and the constraints (\ref{A.23}) satisfy the Dirac bracket superalgebra (\ref{aa2.73}), we need to calculate the Dirac brackets of the supercharges and the Hamiltonian:
\begin{align} \label{A.45}
\{Q_\alpha, \mathit{H} \}_D & = \{Q_1, \mathit{H} \}_D + \{Q_2, \mathit{H} \}_D \:.
\end{align}
To simplify the calculation of Eq. (\ref{A.45}), let us calculate the Poisson bracket of the supercharges and the Hamiltonian.
\begin{align} \label{A.46}
\{Q_1, \mathit{H} \}_P &= \left( \frac{\partial Q_1}{\partial q_i} \frac{\partial \mathit{H}}{\partial p^i} - \frac{\partial Q_1}{\partial p^i} \frac{\partial \mathit{H}}{\partial q_i} \right) 
- \left( \frac{\partial Q_{1}}{\partial \psi_{\alpha}} \frac{\partial \mathit{H}}{\partial \pi^{\alpha}} 
+ \frac{\partial Q_1}{\partial \pi^{\alpha}} \frac{\partial \mathit{H}}{\partial \psi_{\alpha}} \right) \nonumber \\
&=  \omega p \psi_2 - \omega^2 q \psi_1 \:,
\end{align}
\begin{align}\label{A.47}
\{Q_2, \mathit{H} \}_P &= \left( \frac{\partial Q_2}{\partial q_i} \frac{\partial \mathit{H}}{\partial p^i} - \frac{\partial Q_2}{\partial p^i} \frac{\partial \mathit{H}}{\partial q_i} \right) 
- \left( \frac{\partial Q_{2}}{\partial \psi_{\alpha}} \frac{\partial \mathit{H}}{\partial \pi^{\alpha}} 
+ \frac{\partial Q_2}{\partial \pi^{\alpha}} \frac{\partial \mathit{H}}{\partial \psi_{\alpha}} \right) \nonumber \\
&=  - \omega p \psi_1 - \omega^2 q \psi_2 \:,
\end{align}
\begin{align}\label{A.48}
\{\phi_1, \mathit{H} \}_P &= \left( \frac{\partial \phi_1}{\partial q_i} \frac{\partial \mathit{H}}{\partial p^i} - \frac{\partial \phi_1}{\partial p^i} \frac{\partial \mathit{H}}{\partial q_i} \right) 
- \left( \frac{\partial \phi_1}{\partial \psi_{\alpha}} \frac{\partial \mathit{H}}{\partial \pi^{\alpha}} 
+ \frac{\partial \phi_1}{\partial \pi^{\alpha}} \frac{\partial \mathit{H}}{\partial \psi_{\alpha}} \right) \nonumber \\
&=  i \omega \psi_2 \:,
\end{align}
\begin{align} \label{A.49}
\{\phi_2, \mathit{H} \}_P &= \left( \frac{\partial \phi_2}{\partial q_i} \frac{\partial \mathit{H}}{\partial p^i} - \frac{\partial \phi_2}{\partial p^i} \frac{\partial \mathit{H}}{\partial q_i} \right) 
- \left( \frac{\partial \phi_2}{\partial \psi_{\alpha}} \frac{\partial \mathit{H}}{\partial \pi^{\alpha}} 
+ \frac{\partial \phi_2}{\partial \pi^{\alpha}} \frac{\partial \mathit{H}}{\partial \psi_{\alpha}} \right) \nonumber \\
&=  -i \omega \psi_1 \:.
\end{align}

Therefore, using Eqs. ( \ref{A.46},\ref{A.47}, \ref{A.48}, and \ref{A.49}), we calculate the Dirac bracket of the supercharges and the Hamiltonian.
\begin{align} \label{A.50}
\{Q_1, \mathit{H}\}_D & = \{Q_1, \mathit{H}\}_P - \{ Q_1, \phi^A \}_P C_{AB} \{ \phi^B, \mathrm{H} \}_P \nonumber \\
& = \resizebox{.38 \textwidth}{!}  {$\{Q_1, \mathit{H}\}_P - \{ Q_1, \phi^1 \}_P C_{11} \{ \phi^1, \mathit{H} \}_P - \{ Q_1, \phi^2 \}_P C_{22} \{ \phi^2, \mathit{H} \}_P $}\nonumber \\
& =   \omega p \psi_2 - \omega^2 q \psi_1 - \omega p \psi_2 + \omega^2 q \psi_1.
\end{align}
\begin{align} \label{A.51}
\{Q_2, \mathit{H}\}_D & = \{Q_2, \mathit{H}\}_P - \{ Q_2, \phi^A \}_P C_{AB} \{ \phi^B, \mathit{H} \}_P  \nonumber \\
& = \resizebox{.38 \textwidth}{!}  {$\{Q_2, \mathit{H}\}_P - \{ Q_2, \phi^1 \}_P C_{11} \{ \phi^1, \mathit{H} \}_P - \{ Q_2, \phi^2 \}_P C_{22} \{ \phi^2, \mathit{H} \}_P$} \nonumber \\
& = - \omega p \psi_1 - \omega^2 q \psi_2 + \omega^2 q \psi_2 + \omega p \psi_1 \:,
\end{align}
Substitute Eqs. (\ref{A.50}, \ref{A.51})  into equation (\ref{A.45}), thus, we find
\begin{align} \label{A.52}
\{Q_\alpha, \mathit{H} \}_D & = \{Q_1, \mathit{H} \}_D + \{Q_2, \mathit{H} \}_D \nonumber \\
& = \left( \omega p \psi_2 - \omega^2 q \psi_1 - \omega p \psi_2 + \omega^2 q \psi_1 \right) \nonumber \\
& \quad + \left( - \omega p \psi_1 - \omega^2 q \psi_2 + \omega^2 q \psi_2 + \omega p \psi_1 \right) \nonumber \\
&= 0 \:.
\end{align}
It is clear from Eqs. (\ref{A.44} and \ref{A.52}) that the supercharges (\ref{A.24}) together with the Hamiltonian (\ref{A.22}) and the constraints (\ref{A.23}) satisfy the two conditions (\ref{aa2.73}) of the Dirac bracket superalgebra.

\section{The First-Order correction} \label{AppendixB.1}

In this appendix, we use conventional perturbation theory to compute the first-order correction to the ground-state energy of the supersymmetric harmonic oscillator.  Good discussion of the perturbation theory can be found in \cite{goldstein1965classical,griffiths2005introduction,book:17492,sakurai2011modern,shankar2012principles}. In the nondegenerate time-independent perturbation theory, the first-order correction to the energy of the $n^{th}$' state is given by
\begin{equation} \label{B.1}
\mathit{E}_n^{(1)} = \langle \psi_n^0 \vert \mathit{\hat{H}}' \vert \psi_n^0 \rangle.
\end{equation}
We need now to use this formula to compute the first-order correction to the ground-state energy of our system. As we have shown in Section \ref{Sec2.6}, the supersymmetric quantum mechanics Hamiltonian is given in the general form
\begin{equation}
\mathit{\hat{H}} = \frac{1}{2} (\hat{p}^2 + \mathit{W}'^2) \mathbbmtt{I}_2 + \frac{1}{2} \hbar W'' \sigma_3 \:.
\end{equation}  
Let us here consider the superpotential $\mathit{W}(x)$ defined as
\begin{equation}
\mathit{W}= \frac{1}{2} \omega \hat{q}^2 + \mathit{g} \hat{q}^3 \quad \Rightarrow \quad \begin{matrix}  \;\;\; \mathit{W}'  =  \omega \hat{q}+ 3 \mathit{g} \hat{q}^2 \\ \mathit{W}''  = \omega + 6 \mathit{g} \hat{q} \:, \end{matrix} 
\end{equation}
where $\mathit{g}$ is a perturbation. If $\mathit{g}$ is zero, the Hamiltonian $\mathit{\hat{H}}$ reduces to the supersymmetric harmonic oscillator Hamiltonian. Therefore, when $\mathit{g}$ is small, the Hamiltonian $\mathit{\hat{H}}$  can be written as 
\begin{equation}
\mathit{\hat{H}} = \mathit{\hat{H}}^0 + \mathit{\hat{H}}' \:,
\end{equation}
where $\mathit{\hat{H}}^{0}$ is the Hamiltonian of the unperturbed supersymmetric harmonic oscillator and $\mathit{\hat{H}}'$ is the perturbation:
\begin{equation}
\mathit{\hat{H}}'= (\frac{9}{2} \mathit{g}^2 \hat{q}^4 + 3 \omega \mathit{g} \hat{q}^3) \mathbbmtt{I}_2 + 3 \hbar \mathit{g} \hat{q} \sigma_3 \:.
\end{equation}
As shown in Section \ref{Sec2.5}, for the supersymmetric harmonic oscillator, except the ground energy state, all the energy levels degenerate into two energy states. Therefore, if we have an energy level degenerates to the two energy states $\Psi_n=\left( \begin{matrix} \psi_n^a \\ 0 \end{matrix} \right)$ and $\Psi_n= \left( \begin{matrix} 0\\ \psi_n^b \end{matrix} \right)$, we can write
\begin{equation} 
\langle \psi_n \vert \mathit{\hat{H}}' \vert \psi_m \rangle = \Big\langle\left( \begin{matrix} \psi_n^a \\ \psi_n^b \end{matrix} \right) \Big\vert f_1(\hat{q}) \mathbbmtt{I}_2 + f_2(\hat{q}) \sigma_3 \Big\vert \left( \begin{matrix} \psi_m^a \\ \psi_m^b \end{matrix} \right)  \Big\rangle  = 0 \:.
\end{equation} 
If we interest in a ground state of the form $\left(\begin{matrix} 0 \\ \psi_0 \end{matrix}\right)$, the previous equation reduces to
\begin{equation}
\langle \psi_0 \vert \mathit{\hat{H}}' \vert \psi_m \rangle = \langle \psi_0 \vert f_1 (\hat{q}) - f_2 (\hat{q}) \vert \psi_m \rangle  = 0 \:.
\end{equation}
Based on this argument, we can see for this combination that the impact of the Hamiltonian $\mathit{\hat{H}}'$ on the wave function $\Psi_0^{(0)}$ is only due to the component:
\begin{equation} \label{B.8}
\mathcal{\hat{H}}'= \frac{9}{2} \mathit{g}^2 \mathit{q}^4 + 3 \omega \mathit{g} \hat{q}^3 - 3 \hbar \mathit{g} \hat{q} \:.
\end{equation}
Actually, this allows us to compute the first and second corrections to the ground-state energy for this system just using the Hamiltonian $\mathcal{\hat{H}}'$. As well, we can use the nondegenerate time-independent perturbation theory to calculate the corrections to the ground-state energy of our harmonic oscillator system even if it has degenerate energy levels.

Thus, taking into account the supersymmetric harmonic oscillator ground-state wave function in the form (\ref{f2.142}) and the Hamiltonian (\ref{B.8}), then, we can now use the formula (\ref{B.1}) to compute the first-order correction to the energy of the ground state as follows:
\begin{align}
\mathit{E}_0^{(1)} & = \langle \psi_0^0 \vert \mathcal{\hat{H}}' \vert \psi_0^{(0)} \rangle \nonumber \\
& = \langle \psi_0^{(0)} \vert (\frac{9}{2} \mathit{g}^2 \hat{q}^4 + 3\omega \mathit{g} \hat{q}^3) \mathbbmtt{I}_2 + 3 \hbar \mathit{q} \hat{q} \sigma_3 \vert \psi_0^{(0)} \rangle  \nonumber  \\
&= \int_{-\infty}^{+\infty} \sqrt{\frac{\omega}{\pi \hbar}} e^{- \frac{\omega \hat{q}^2}{\hbar}} \left( \frac{9}{2} \mathit{g}^2 \hat{q}^4 + 3 \omega \mathit{g} \hat{q}^3 - 3 \hbar \mathit{g} \hat{q} \right) d q \nonumber \\
&= \frac{9 \mathit{g}^2}{2} \sqrt{\frac{\omega}{\pi \hbar}} \int_{-\infty}^{+\infty} \hat{q}^4 e^{- \frac{\omega \hat{q}^2}{\hbar}} dq + 3 \omega \mathit{g} \sqrt{\frac{\omega}{\pi \hbar}} \int_{-\infty}^{+\infty} \hat{q}^3 e^{- \frac{\omega \hat{q}^2}{\hbar}} dq  \nonumber \\
&{\quad}- 3 \hbar \mathit{g} \sqrt{\frac{\omega}{\pi \hbar}} \int_{-\infty}^{+\infty} \hat{q} e^{- \frac{\omega \hat{q}^2}{\hbar}} dq \:.
\end{align}
Both the second term and the third term of the previous integration vanish since the integral functions $\hat{q}^3 \exp(- \omega \hat{q}^2/\hbar)$ and $\hat{q} \exp(- \omega \hat{q}^2/ \hbar)$ are odd functions and there integrations from $- \infty$ to $+ \infty$ are equal to zero. Therefore the previous integration reduces to
\begin{align}\label{B.10}
\mathit{E}_0^{(1)} &=  \frac{9 g^2}{2} \sqrt{\frac{\omega}{\pi \hbar}} \int_{-\infty}^{+\infty} \hat{q}^4 e^{- \frac{\omega \hat{q}^2}{\hbar}} dq = 9 \mathit{g}^2 \sqrt{\frac{\omega}{\pi \hbar}} \int_{0}^{+\infty} e^{- \frac{\omega \hat{q}^2}{\hbar}} \hat{q}^2 dq \nonumber \\ 
&= 9 \mathit{q}^2 \sqrt{\frac{\omega}{\pi \hbar}} \times \frac{3}{8} \sqrt{\frac{\pi \hbar^5}{\omega^5}} = \frac{27}{8}\frac{\hbar^2}{\omega^2} \mathit{g}^2 \simeq \mathcal{O}(g^2) \:.
\end{align}
The last equation implies that the first-order correction to the ground-state energy reduces to zero. Notice that to solve the integration which appeared in the last equation, we have used the integral formula
\begin{equation}
\int_{0}^{\infty}  x^{2n} \exp(- a x^2)dx = \frac{1 \cdot 3 \cdots(2n-1)}{2^{n+1}} \sqrt{\frac{\pi}{a^{2n+1}}} \:.
\end{equation}

\section{The Second-Order Correction } \label{AppendixB.2}

As described in \cite{goldstein1965classical,griffiths2005introduction,book:17492,sakurai2011modern,shankar2012principles}, the second-order correction to the energy in the time-independent nondegenerate perturbation theory is given by the form
\begin{equation} \label{B.12}
\mathit{E}_0^{(2)} = \sum_{m \neq n} \frac{\vert \langle \psi_n^{(0)} \vert \mathcal{\hat{H}}' \vert \psi_m^{(0)} \rangle \vert^2}{\mathit{E}_n^{(0)} - \mathit{E}_m^{(0)}}\:.
\end{equation}
Furthermore, the harmonic oscillator wave function is orthogonal polynomial, and the harmonic oscillator wave function of degree $n$ can be obtained by the following recursion formula:
\begin{equation} \label{B.13}
\psi_n^{(0)} (\hat{q})= \frac{1}{\sqrt{2^n} n!} \left( \frac{\omega}{\pi \hbar} \right)^{\frac{1}{4}} e^{- \frac{\omega \hat{q}^2}{2 \hbar}} \mathit{\hat{H}}_n \left(\sqrt{\frac{\omega}{\hbar}} \hat{q} \right)\:, 
\end{equation} 
where $\mathit{\hat{H}}_n$ is the Hermite Polynomial of degree $n$. Also, it is helpful to consider the recursion relation of the Hermite Polynomials:
\begin{equation} \label{B.14}
\hat{q} \mathit{\hat{H}}_n = \frac{1}{2} \mathit{\hat{H}}_{n+1} + n \mathit{\hat{H}}_{n-1}\:.
\end{equation}
Moreover, for two wave functions $\psi_n$ and $\psi_m$, we have the relation
\begin{equation} \label{B.15}
\langle \psi_n \vert \psi_m \rangle = \delta_{nm}\:.
\end{equation} 
Eqs. (\ref{B.13}, \ref{B.14}, and \ref{B.15}) imply that
\begin{equation}
\langle \psi_0 \vert q \vert \psi_m \rangle = \frac{1}{2} C_{11} \delta_{0(m+1)} + C_{12} n \delta_{0(m-1)} = C_{12} n \delta_{1m}\:.
\end{equation}
This argument leads us to know that all the terms of the summation with $m>4$ in Eq. (\ref{B.12}) vanish. So, to compute the second-order correction to the energy of the ground state using Eq. (\ref{B.12}), we only need to consider the nonzero terms with $m \leq 4$.

Now, we are able to compute the first term with $m=1$. First considering the ground and the first excited state wave functions:
\begin{equation}
\psi_0^{(0)} =(\frac{\omega}{\pi \hbar})^{\frac{1}{4}} e^{- \frac{\omega q^2}{2 \hbar}}, \quad \texttt{and} \quad
\psi_1^{(0)} = (\frac{4\omega^3}{\pi \hbar^3})^{\frac{1}{4}} q e^{- \frac{\omega q^2}{2 \hbar}}\:.
\end{equation} 
Then,
\begin{align}\label{B.18}
\MoveEqLeft \resizebox{.48 \textwidth}{!}  {$ \frac{\vert \langle \psi_0^{(0)} \vert 3\omega \mathit{g} \hat{q}^3 -3 \hbar \mathit{g} \hat{q}  \vert \psi_1^{(0)} \rangle \vert^2}{\mathit{E}_0^{(0)} - \mathit{E}_1^{(0)}} = \frac{\Big\vert 3\omega \mathit{g} \langle \psi_0^{(0)} \vert \hat{q}^3 \vert \psi_1^{(0)} \rangle - 3 \hbar \mathit{g} \langle \psi_0^{(0)} \vert \hat{q} \vert \psi_1^{(0)} \rangle \Big\vert^2}{ - \hbar \omega} $} \nonumber \\ 
& = \resizebox{.42 \textwidth}{!}  {$ \frac{-1}{\hbar \omega} \left[ 6 \omega \hat{g} \sqrt{\frac{2 \omega^2}{\pi \hbar^2}} \int_0^{\infty} \hat{q}^4 e^{\frac{-\omega \hat{q}^2}{\hbar}} dq - 6 \hbar \mathit{g} \sqrt{\frac{2 \omega^2}{\pi \hbar^2}} \int_0^{\infty} \hat{q}^2 e^{\frac{-\omega \hat{q}^2}{\hbar}} dq \right]^2 $} \nonumber \\ 
& = \frac{-\mathit{g}^2}{\hbar \omega} \left[\frac{9}{4} \sqrt{\frac{2 \hbar^3}{\omega}} - \frac{3}{2} \sqrt{\frac{2 \hbar^3}{\omega}} \right]^2 = - \frac{9}{8}\frac{\hbar^2}{\omega^2} \mathit{g}^2\:.
\end{align}
Similarly, we can compute the second term with $m=2$. The second excited state wave functions is given by the form
\begin{equation}
\psi_2^{(0)}= \left( \frac{\omega}{ 4 \pi \hbar}\right)^{\frac{1}{4}}\left( \frac{ 2 \omega}{\hbar} \hat{q}^2 - 1 \right) e^{- \frac{\omega q^2}{2 \hbar}}\:.
\end{equation} 
Then,
\begin{align}\label{B.20}
\resizebox{.12 \textwidth}{!}  {$ \frac{\vert \langle \psi_0^{(0)} \vert \frac{9}{2} \mathit{g}^2 \hat{q}^4  \vert \psi_2^{(0)} \rangle \vert^2}{\mathit{E}_0^{(0)} - \mathit{E}_2^{(0)}} $}
& = \resizebox{.33 \textwidth}{!}  {$ \frac{\Big\vert \Big\langle (\frac{\omega}{\pi \hbar})^{\frac{1}{4}} e^{- \frac{\omega \hat{q}^2}{2 \hbar}} \Big\vert \frac{9}{2} \mathit{g}^2 \hat{q}^4 \Big\vert \left(\frac{\omega}{4 \pi \hbar}\right)^{\frac{1}{4}}\left( \frac{2\omega}{\hbar} \hat{q}^2 -1\right) e^{- \frac{\omega \hat{q}^2}{2 \hbar}} \Big\rangle \Big\vert^2}{0 - 2 \hbar \omega} $} \nonumber \\
&= \resizebox{.33 \textwidth}{!}  {$\frac{-81 \mathit{g}^4}{16 \pi \hbar^2} \left[ \Big\vert \frac{4\omega}{\hbar} \int_{0}^{+\infty} \hat{q}^6 e^{-\frac{\omega \hat{q}^2}{\hbar}} dq -2 \int_{0}^{+\infty} \hat{q}^4 e^{-\frac{\omega \hat{q}^2}{\hbar}} dq \Big\vert^2 \right]$} \nonumber \\
&= \resizebox{.33 \textwidth}{!}  {$ \frac{-81 \mathit{g}^4}{16 \pi \hbar^2} \left( \frac{15}{4} \sqrt{\frac{\pi \hbar^5}{\omega^5}} - \frac{3}{4} \sqrt{\frac{\pi \hbar^5}{\omega^5}}  \right)^2 = - \frac{729}{16} \frac{\hbar^3}{\omega^5} \mathit{g}^4\:.$}
\end{align}
Furthermore, the same way we can compute the second term with $m=3$. The  third excited state wave functions take the form
\begin{equation}
\psi_3^{(0)}= (\frac{\omega^3}{9 \pi \hbar^3})^{\frac{1}{4}}(2 \frac{\omega}{\hbar} \hat{q}^3-3 \hat{q}) e^{- \frac{\omega \hat{q}^2}{2 \hbar}}\:.
\end{equation} 
Then,
\begin{align}\label{B.22}
\resizebox{.12 \textwidth}{!}  {$ \frac{\vert \langle \psi_0^{(0)} \vert 3 \omega \mathit{g} \hat{q}^3  \vert \psi_3^{(0)} \rangle \vert^2}{\mathit{E}_0^{(0)} - \mathit{E}_3^{(0)}}$}
& = \resizebox{.33 \textwidth}{!}  {$\frac{\Big\vert \Big\langle (\frac{\omega}{\pi \hbar})^{\frac{1}{4}} e^{- \frac{\omega q^2}{2 \hbar}} \Big\vert 3 \omega gq^3  \Big\vert (\frac{\omega^3}{9 \pi \hbar^3})^{\frac{1}{4}}(2 \frac{\omega}{\hbar} \hat{q}^3-3 \hat{q}) e^{- \frac{\omega \hat{q}^2}{2 \hbar}} \Big\rangle \Big\vert^2}{0 - 3 \hbar \omega}$} \nonumber \\
& = \resizebox{.33 \textwidth}{!}  {$\frac{- \mathit{g}^2 \omega^3}{\pi \hbar^3} \left[ \Big\vert \Big\langle e^{- \frac{\omega \hat{q}^2}{2 \hbar}} \Big\vert \hat{q}^3 \Big\vert (2 \frac{\omega}{\hbar} \hat{q}^3-3 \hat{q}) e^{- \frac{\omega \hat{q}^2}{2 \hbar}} \Big\rangle \Big\vert^2 \right]$} \nonumber \\ 
& = \resizebox{.33 \textwidth}{!}  {$\frac{- \mathit{g}^2 \omega^3}{\pi \hbar^3} \left[ \Big\vert \frac{2\omega}{\hbar} \int_{-\infty}^{+\infty} \hat{q}^6 e^{- \frac{\omega \hat{q}^2}{\hbar}} dq - 3 \int_{-\infty}^{+\infty} \hat{q}^4 e^{- \frac{\omega \hat{q}^2}{\hbar}} dq \Big\vert^2 \right]$} \nonumber \\
&= \resizebox{.33 \textwidth}{!}  {$\frac{- \mathit{g}^2 \omega^3}{\pi \hbar^3} \left[ \Big\vert \frac{4\omega}{\hbar} \int_{0}^{+\infty} \hat{q}^6 e^{- \frac{\omega \hat{q}^2}{\hbar}} dq - 6 \int_{0}^{+\infty} \hat{q}^4 e^{- \frac{\omega \hat{q}^2}{\hbar}} dq \Big\vert^2 \right]$} \nonumber \\
& = \resizebox{.33 \textwidth}{!}  {$\frac{- \mathit{g}^2 \omega^3}{\pi \hbar^3} \left[ \Big\vert \frac{15}{4} \sqrt{\frac{\pi \hbar^5}{\omega^5}} - \frac{9}{4} \sqrt{\frac{\pi \hbar^5}{\omega^3}} \Big\vert^2 \right] = \frac{- \mathit{g}^2 \omega^3}{\pi \hbar^3} \left(\frac{3}{2} \sqrt{\frac{\pi \hbar^5}{\omega^3}} \right)^2$}  \nonumber \\
& = - \frac{9}{4}\frac{\hbar^2}{\omega^2} \mathit{g}^2 \:.
\end{align}
Also, similarly, we can compute the term with $m=4$. The fourth excited state wave function is
\begin{equation}
\psi_4^{(0)}= \left(\frac{\omega}{576 \pi \hbar}\right)^{\frac{1}{4}}\left( \frac{4\omega^2}{\hbar^2} \hat{q}^4 -\frac{12\omega}{\hbar} \hat{q}^2 +3\right) e^{- \frac{\omega \hat{q}^2}{2 \hbar}}\:.
\end{equation} 
Then,
\begin{align}\label{B.24}
\MoveEqLeft \resizebox{.12 \textwidth}{!}  {$\frac{\vert \langle \psi_0^{(0)} \vert \frac{9}{2} \mathit{g}^2 \hat{q}^4  \vert \psi_4^{(0)} \rangle \vert^2}{\mathit{E}_0^{(0)} - \mathit{E}_4^{(0)}}$}
 = \resizebox{.33 \textwidth}{!}  {$ \frac{\Big\vert \Big\langle (\frac{\omega}{\pi \hbar})^{\frac{1}{4}} e^{- \frac{\omega \hat{q}^2}{2 \hbar}} \Big\vert \frac{9}{2} \mathit{g}^2 \hat{q}^4 \Big\vert \left(\frac{\omega}{576 \pi \hbar}\right)^{\frac{1}{4}}\left( \frac{4\omega^2}{\hbar^2} \hat{q}^4 -\frac{12\omega}{\hbar} \hat{q}^2 +3\right) e^{- \frac{\omega \hat{q}^2}{2 \hbar}} \Big\rangle \Big\vert^2}{0 - 4 \hbar \omega} $} \nonumber \\
& = \resizebox{.40 \textwidth}{!}  {$\frac{-27 \mathit{g}^4}{128 \pi \hbar^2} \left[ \Big\vert \Big\langle e^{-\frac{\omega \hat{q}^2}{2\hbar}} \Big\vert \hat{q}^4 \Big\vert \left( \frac{4\omega^2}{\hbar^2} \hat{q}^4 -\frac{12\omega}{\hbar} \hat{q}^2 +3\right) e^{- \frac{\omega \hat{q}^2}{2 \hbar}} \Big\rangle \Big\vert^2 \right]$} \nonumber \\ 
& = \resizebox{.40 \textwidth}{!}  {$ \frac{-27 \mathit{g}^4}{128 \pi \hbar^2} \left[ \Big\vert \frac{4\omega^2}{\hbar^2} \int_{-\infty}^{+\infty} \hat{q}^8 e^{-\frac{\omega \hat{q}^2}{\hbar}} dq -\frac{12\omega}{\hbar} \int_{-\infty}^{+\infty} \hat{q}^6 e^{-\frac{\omega \hat{q}^2}{\hbar}} dq +3 \int_{-\infty}^{+\infty} \hat{q}^4 e^{-\frac{\omega \hat{q}^2}{\hbar}} dq \Big\vert^2 \right]$} \nonumber \\
& =\resizebox{.40 \textwidth}{!}  {$ \frac{-27 \mathit{g}^4}{32 \pi \hbar^2} \left[ \Big\vert \frac{4\omega^2}{\hbar^2} \int_{0}^{+\infty} \hat{q}^8 e^{-\frac{\omega \hat{q}^2}{\hbar}} dq -\frac{12\omega}{\hbar} \int_{0}^{+\infty} \hat{q}^6 e^{-\frac{\omega \hat{q}^2}{\hbar}} dq +3 \int_{0}^{+\infty} \hat{q}^4 e^{-\frac{\omega \hat{q}^2}{\hbar}} dq \Big\vert^2 \right]$} \nonumber \\
& = \resizebox{.40 \textwidth}{!}  {$\frac{-27 \mathit{g}^4}{32 \pi \hbar^2} \left[ \Big\vert\frac{105}{8} \sqrt{\frac{\pi \hbar^5}{\omega^5}} - \frac{45}{4} \sqrt{\frac{\pi \hbar^5}{\omega^5}} + \frac{9}{8} \sqrt{\frac{\pi \hbar^5}{\omega^5}} \Big\vert^2 \right] = \frac{-27 \mathit{g}^4}{32 \pi \hbar^2} \left[ \Big\vert 3 \sqrt{\frac{\pi \hbar^5}{\omega^5}} \Big\vert^2 \right]$} \nonumber \\
& = - \frac{243}{32} \frac{\hbar^3}{\omega^5} \mathit{g}^4\:.
\end{align}
Using Eqs. (\ref{B.18}, \ref{B.20}, \ref{B.22}, and \ref{B.24}), we are able to compute the second-order correction to the ground-state energy for the supersymmetric harmonic oscillator as follows:
\begin{align} \label{B.25}
\mathit{E}_2^{(0)} & = \frac{\vert \langle \psi_0^{(0)} \vert - 3 \hbar \mathit{g} \hat{q}  \vert \psi_1^{(0)} \rangle \vert^2}{\mathit{E}_0^{(0)} - \mathit{E}_1^{(0)}} + \frac{\vert \langle \psi_0^{(0)} \vert \frac{9}{2} \mathit{g}^2 \hat{q}^4  \vert \psi_2^{(0)} \rangle \vert^2}{\mathit{E}_0^{(0)} - \mathit{E}_2^{(0)}}  \nonumber \\
& \quad + \frac{\vert \langle \psi_0^{(0)} \vert 3 \omega \mathit{g}  \hat{q}^3  \vert \psi_3^{(0)} \rangle \vert^2}{\mathit{E}_0^{(0)} - \mathit{E}_3^{(0)}} +  \frac{\vert \langle \psi_0^{(0)} \vert \frac{9}{2} \mathit{g}^2 \hat{q}^4  \vert \psi_4^{(0)} \rangle \vert^2}{\mathit{E}_0^{(0)} - \mathit{E}_4^{(0)}} \nonumber \\
& = - \frac{9}{8}\frac{\hbar^2}{\omega^2} \mathit{g}^2 - \frac{729}{16} \frac{\hbar^3}{\omega^5} \mathit{g}^4 - \frac{9}{4}\frac{\hbar^2}{\omega^2} \mathit{g}^2  - \frac{243}{32} \frac{\hbar^3}{\omega^5} \mathit{g}^4 \nonumber \\
&= - \frac{27}{8}\frac{\hbar^2}{\omega^2} \mathit{g}^2  - \frac{1701}{32} \frac{\hbar^3}{\omega^5} \mathit{g}^4\:.
\end{align}    
Finally, from Eqs. (\ref{B.10}) and (\ref{B.25}), we realize that up to the second order the energy corrections to the ground-state energy of the supersymmetric quantum mechanical harmonic oscillator vanish. This is concluded as
\begin{align}
\mathcal{O}(g)&=0, \nonumber \\
\mathcal{O}(g^2) &= \frac{27}{8}\frac{\hbar^2}{\omega^2} - \frac{27}{8}\frac{\hbar^2}{\omega^2}=0\:.
\end{align}
  
\bibliographystyle{unsrt}
\bibliography{references}

\end{document}